\pdfoutput=1

\documentclass[prx,letterpaper,nobalancelastpage,superscriptaddress,nofootinbib]{revtex4-2}

\usepackage{graphicx}
\usepackage{amsmath}
\usepackage{amssymb}
\usepackage[english]{babel}
\usepackage{color}
\usepackage[version=4]{mhchem}
\usepackage{dsfont}
\usepackage{placeins}
\usepackage{mathtools}

\usepackage{caption}

\DeclareCaptionLabelSeparator{bar}{ \textbf{\textbar}~}
\usepackage{enumitem}
\setlist{nolistsep}

\newcommand{\YBCO}{$\mathrm{YBa_{2}Cu_{3}O_{7-\delta}}$}
\newcommand{\YXY}{$\mathrm{(Y_{1-x-y}Ca_{x}Ce_{y})Ba_{2}Cu_{3}O_{7-\delta}}$}

\newcommand{\Caltech}{Engineering \& Applied Science, California Institute of Technology, Pasadena, CA 91125, USA}

\newcommand{\CAN}{CAN SUPERCONDUCTORS, s.r.o., Ringhofferova 66, 251 68 Kamenice, Czech Republic}

\newcommand{\Towson}{Department of Physics, Astronomy \& Geosciences, Towson University, 8000 York Road, Towson, MD 21252, USA}

\newcommand{\Tomas}{Department of Inorganic Chemistry, University of Chemistry and Technology Prague, Technick\'a 5, 166 28, Prague 6, Czech Republic}

\begin{document}

\title{Potential Major Improvement in Superconductors for High-Field Magnets}

\author{Jamil Tahir-Kheli}
\affiliation{\Caltech}

\author{Tom\'a\v s Hl\'asek}
\affiliation{\CAN}
\affiliation{\Tomas}

\author{Michal Lojka}
\affiliation{\Tomas}
\affiliation{\CAN}

\author{Michael S. Osofsky}
\affiliation{\Towson}

\author{Carver A. Mead}
\affiliation{\Caltech}

\maketitle

\textbf{
Fusion reactors are limited by the magnetic field available to confine their plasma. The commercial fusion industry uses the larger magnetic field and higher operating temperature of the cuprate superconductor
$\mathbf{YBa_{2}Cu_{3}O_{7-\delta}}$ (YBCO) in order to confine their plasma into a dense volume.
A superconductor is a macroscopic quantum state that is protected from the metallic (resistive) state by an energy gap.
Unfortunately, YBCO has an anisotropic gap, known as D-wave because it has the shape of a $\mathbf{d_{x^2-y^2}}$ chemical orbital. This D-wave gap means that poly-crystalline wire cannot be made because a few degree misalignment between grains in the wire leads to a drastic loss in its supercurrent carrying ability, and thereby its magnetic field limit. The superconductor industry has responded by growing nearly-single-crystal superconducting YBCO films on carefully prepared substrate tapes kilometers in length. Heroic development programs have made such tapes commercially available, but they are very expensive and delicate.
MRI magnet superconductors, such as $\mathbf{NbTi}$ and $\mathbf{Nb_{3}Sn}$, are formed into poly-crystalline wires because they have an isotropic gap in the shape of an s chemical orbital (called S-wave) that makes them insensitive to grain misalignment. However, these materials are limited to lower magnetic fields and liquid-He temperatures.
Here, we modified YBCO by doping the Y site with Ca and Ce atoms to form
$\mathbf{(Y_{1-x-y}Ca_{x}Ce_{y})Ba_{2}Cu_{3}O_{7-\delta}}$,
and show evidence that it changes to an S-wave gap. Its superconducting transition temperature, $\mathbf{T_c}$, of $\mathbf{\sim 70K}$, while lower than that of D-wave YBCO at $\mathbf{\sim 90K}$, is easily maintained using common, economic cryogenic equipment.}

In the popular press, $\mathrm{T_c}$ is considered the most important figure-of-merit of a superconductor. This line of thinking assumes that a higher $\mathrm{T_c}$ always leads to better technology because the main obstacle is the cooling cost. In fact, the maximum sustainable magnetic field at the operating temperature and the cost and reliability of making kilometers of wire are far more important for current and near-future technologies~\cite{Larbelestier2001}. Since YBCO can sustain higher magnetic fields at higher temperatures than $\mathrm{NbTi}$ or $\mathrm{Nb_3Sn}$, the biggest problem with YBCO is making long wires. Thus a natural question to ask is, ``Can we lower the $\mathrm{T_c}$ of D-wave YBCO in order to expose an S-wave YBCO gap phase that will intrinsically perform better when made into poly-crystalline wires?"

It is known~\cite{Hirschfeld1993} that non-magnetic impurities smear out the superconducting gap anisotropy. For a D-wave superconductor, this gap smearing reduces its gap anisotropy, and thereby its $\mathrm{T_c}$~\cite{Abrikosov1961}. Smearing an isotropic gap has no affect on the overall gap isotropy. Thus, non-magnetic impurities do not affect the $\mathrm{T_c}$ of S-wave superconductors~\cite{Anderson1959}.
Hence, the plan in this paper is to add sufficient non-magnetic impurities to YBCO such that the D-wave $\mathrm{T_c}$ becomes smaller than the S-wave $\mathrm{T_c}$.

\begin{figure}[tbp]
\centering \includegraphics[width=0.50\textwidth]{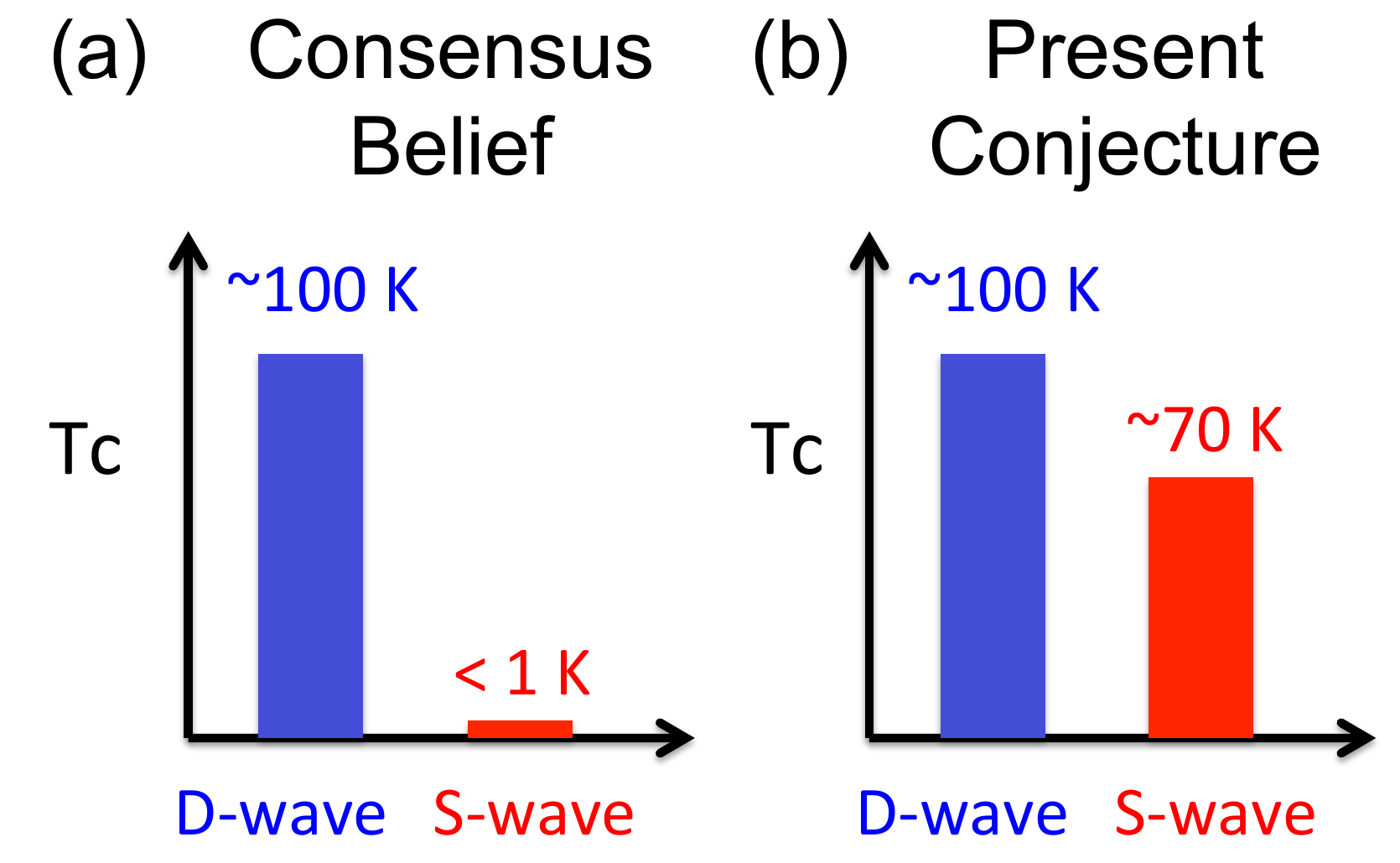}
\caption{
\textbf{Is a technologically useful YBCO S-wave superconductor hiding just below the difficult to manage D-wave YBCO?}
(a) shows the consensus belief that any S-wave YBCO phase would have its $\mathrm{T_c}$ below $1\ \mathrm{K}$, and thus would not be a useful technology. (b) shows our conjecture that an S-wave phase in YBCO exists at $\sim 70\ \mathrm{K}$. If true, unearthing this phase would have huge implications for high-magnetic-field applications. Here, we tried to ``push down" the D-wave phase $\mathrm{T_c}$, while not affecting the S-wave phase $\mathrm{T_c}$. Our approach is to add non-magnetic Ca and Ce atoms to
\YBCO\ to form \YXY\ [hereinafter, denoted by (X,Y)]. Our experimental results are shown in Figures~\ref{tcplot}, \ref{pcar}, and \ref{lambda}. Table~\ref{list} summarizes the results. Taken together, they suggest that an S-wave YBCO phase does exist at $\sim 70\ \mathrm{K}$.
}
\label{tcdvs}
\end{figure}

Figure~\ref{tcdvs} shows the two possible scenarios for the S-wave $\mathrm{T_c}$ of YBCO relative to its D-wave $\mathrm{T_c}$. Figure~\ref{tcdvs}a is the consensus expectation that the S-wave $\mathrm{T_c}$ is so small such that, if ever uncovered, it would be practically useless due to cooling costs. Figure~\ref{tcdvs}b is our conjecture for where the S-wave $\mathrm{T_c}$ is located~\cite{JTK2017}. If Figure~\ref{tcdvs}b is true, then there is enormous value to current magnet technologies in uncovering this S-wave phase of YBCO.

To decide between Figure~\ref{tcdvs}a and Figure~\ref{tcdvs}b, three experiments were performed on YBCO samples with different concentrations of non-magnetic impurities. Each experiment was chosen because it analyzed a distinct fundamental physical characteristic of the superconducting gap.

\FloatBarrier

\FloatBarrier

\section{Three Experiments to distinguish between D-wave and S-wave gap symmetries}

The superconducting gap is represented by a complex number that is a function of direction with respect to the crystal axes of YBCO. It has magnitude and phase in every direction. The gap is invariant to an overall phase change. The $\mathrm{T_c}$ is proportional to the maximum magnitude of the gap. For S-wave and D-wave gap symmetries, the gap function can be taken to be real in every direction. An S-wave gap is positive, while a D-wave gap is positive and negative with a zero in-between. The D-wave gap in YBCO has zeros, or nodes, along directions $\mathrm{45^o}$ from the axes along the Cu-O bond directions in the $\mathrm{CuO_2}$ planes and opposite signs along the two perpendicular Cu-O bond directions.

\begin{figure}[!hb]
\centering \includegraphics[width=100mm]{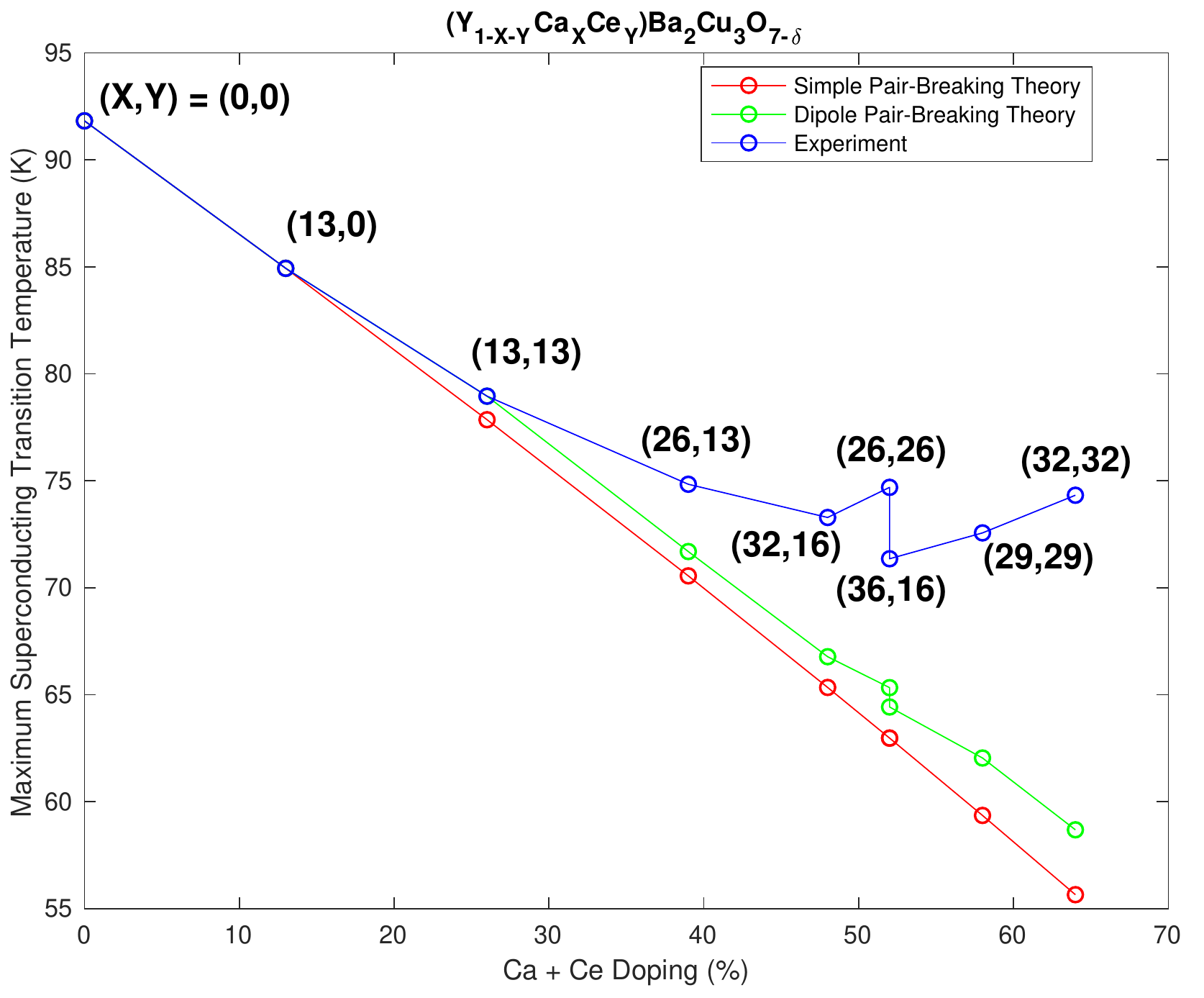}
\caption{
\textbf{Evolution of the maximum superconducting $\mathbf{T_{c,max}}$ {\it vs} Ca and Ce counter-doping in YBCO compared to D-wave theory predictions.}
Blue points are $\mathrm{T_{c,max}}$ which initially drops with counter-doping and then ``saturates" at $\sim 72\ \mathrm{K}$. Saturation of $\mathrm{T_{c,max}}$ suggests the known D-wave of pure YBCO has changed to S-wave because the $\mathrm{T_c}$ of S-wave superconductors is only weakly dependent on non-magnetic counter-doping~\cite{Anderson1959} (Anderson's Theorem). $\mathrm{T_{c,max}}$ is the maximum $\mathrm{T_c}$ for each $\mathrm{(X,Y)}$ after low-temperature annealing that changes the Oxygen content of the sample.
Red and green points are the predicted $\mathrm{T_{c,max}}$ results. Red points assume simple pair-breaking, where Ca and Ce atoms lead to identical pair-breaking strengths. Green points assume that Ca and Ce atoms are close together in the material because Ca and Ce have $\mathrm{+1}$ and $\mathrm{-1}$ charges relative to Y, respectively (hence, the name ``dipole pair-breaking"). Details are given in the Supplement~\cite{Supplement}.
The red and green points do not explain the experiment (blue points) suggesting that highly counter-doped YBCO is an S-wave superconductor.}
\label{tcplot}
\end{figure}

As described above, the $\mathrm{T_c}$ of D-wave YBCO should fall with increasing non-magneic impurities while the $\mathrm{T_c}$ of an S-wave YBCO phase should remain approximately constant with varying impurity concentrations. The first experiment measures the $\mathrm{T_c}$ evolution with impurity concentration. The results of the experiment are shown in Figure~\ref{tcplot} with additional details of the experiment below.

The second experiment looks for the existence of a sign change in the gap in order to distinguish a D-wave gap from and S-wave gap. A D-wave gap leads to a zero-bias conductance peak (ZBCP) in Point-Contact-Andreev-Reflection (PCAR) tunneling current, while an S-wave gap has no ZBCP. The results of this experiment are shown in Figure~\ref{pcar} with additional details below.

The third experiment searches for nodes in the superconducting gap by measuring the evolution of the superconducting penetration depth, $\mathrm{\lambda}$ as a function of temperature, T, at low-temperatures. A D-wave gap has nodes and will lead to $\mathrm{\lambda\sim T}$, and an S-wave gap leads to $\mathrm{\lambda\sim T^2}$ because it has no nodes~\cite{Tinkham2004}. The results of this experiment are shown in Figure~\ref{lambda} with additional details below.


The non-magnetic impurities used in this paper are Ca atoms that substitute at the Y sites in a $+2$ oxidation state, and Ce atom that substitute at the Y site in a $+4$ oxidation state. Both of these atoms are non-magnetic. Since the oxidation state of Y in YBCO is $+3$, Ca and Ce have a $-1$ and $+1$ charge relative to the Y atoms, respectively.
Also, the ionic radii of Ca (+2), Ce (+4), and Y (+3) are very close. They are $\mathrm{1.00\ \AA}$, $\mathrm{1.11\ \AA}$, and $\mathrm{1.02\ \AA}$, respectively. Thus Ca and Ce atoms do not strain YBCO.
This Ca and Ce charge ``counter-doping" has the benefit of permitting very large amounts of non-magnetic dopants to substitute at the Y site.
Large doping is desirable since we want to push the D-wave $\mathrm{T_c}$ down as far as possible in order to increase our chances of unearthing the S-wave gap phase.

There were many other potential non-magnetic atoms. We chose Ca and Ce because they both substitute at the Y site of YBCO, Ca-doped YBCO is a superconductor~\cite{Williams1998,Naqib2003}, and Ce-doped YBCO is also a superconductor~\cite{Jin1991}. For both Ca and Ce doping, the $\mathrm{T_c}$ is only modestly lower than pure YBCO.

Samples of \YXY\ [hereinafter, denoted by (X,Y)] for (X,Y) = (0.0, 0.0), (0.13, 0.0), (0.13, 0.13), (0.26, 0.13), (0.32, 0.16), (0.36, 0.16), (0.26, 0.26), (0.29, 0.29), and (0.32, 0.32) were synthesized. Since pure YBCO, $\mathrm{(0,0)}$ in our notation, has a superconducting $\mathrm{T_c}$ dome that rises from zero, peaks, and then decreases as the number of Oxygen atoms in the CuO chains increases, we expect that all (X,Y) samples will have similar $\mathrm{T_c}$ domes~\cite{Tallon1995}.

Many low-temperature anneals were done on each sample to change the Oxygen content and thereby obtain the maximum $\mathrm{T_c}$, ($\mathrm{T_{c,max}}$) value for each (X,Y) for all samples reported in this paper. Extended Data Figures~\ref{xrd}, \ref{bsesem2626}, \ref{edx2626}, and Extended Data Tables~\ref{rietveld} and \ref{phasecomp} are materials characterization data that show the samples are single-phase with the stated composition.

\FloatBarrier

\FloatBarrier

\subsection{Evolution of the maximum $\mathrm T_c$ with impurity concentration} 

Figure~\ref{tcplot} shows the maximum superconducting $\mathrm{T_{c,max}}$ as a function of (X,Y) counter-doping. It shows that $\mathrm{T_{c,max}}$ initially falls, as expected for a D-wave superconductor, and then ``saturates" at higher doping. The red and green plots are the results of two different theoretical models, using Abrikosov-Gorkov theory~\cite{Hirschfeld1993,Abrikosov1961}, for the drop in $\mathrm{T_c}$ versus counter-doping if all the samples remained D-wave~\cite{Supplement}. The saturation of the measured $\mathrm{T_{c,max}}$ (blue points) is not compatible with D-wave superconductivity predictions and suggests that highly counter-doped YBCO is an S-wave superconductor.

A very important detail of this experiment is that the maximum $\mathrm{T_c}$ for each counter-doping was used. Tallon et al.~\cite{Presland1991,Obertelli1992} showed that the $\mathrm{T_c}$ dome of cuprate superconductors corresponds to the hole doping in the $\mathrm{CuO_2}$ planes and that this hole doping leads to a unique room-temperature thermopower. From these two relations, $\mathrm{T_{c,max}}$ is predicted to occur when the room-temperature thermopower is $\mathrm{\approx +2\ \mu V/K}$. For all (X,Y) values, this relation was found to be true. We conclude that for each (X,Y), the number of holes in the $\mathrm{CuO_2}$ planes is the same when the transition temperature is maximum. Therefore, any change in $\mathrm{T_{c,max}}$ as a function of (X,Y) is not due to changes in the Fermi level, Fermi surface, or hole doping in the $\mathrm{CuO_2}$ planes.

\subsection{Evolution of a zero-bias-conductance peak in Point-Contact-Andreev-Reflection}

\begin{figure}[!hb]
\includegraphics[width=170mm]{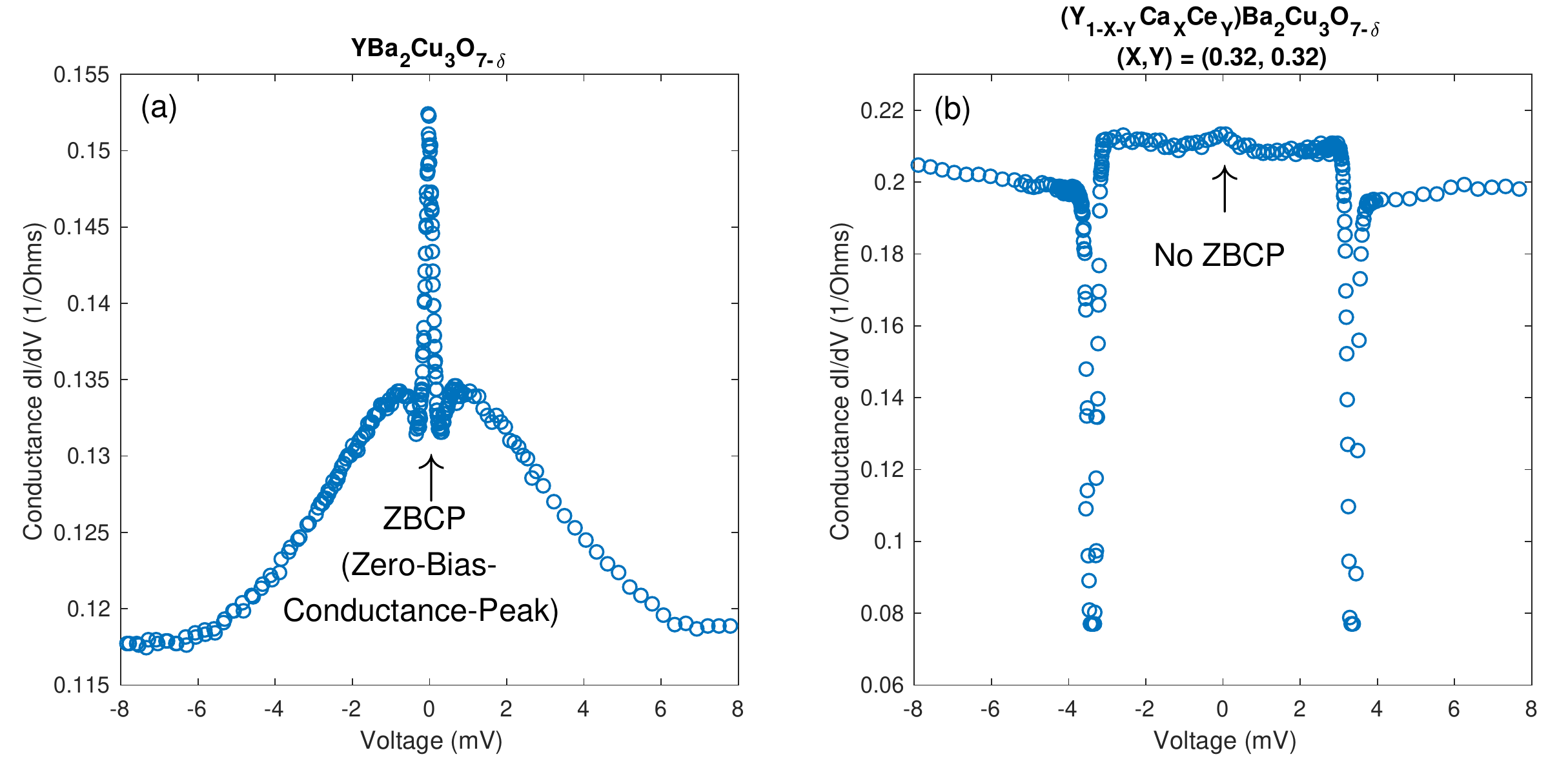}
\caption{
\textbf{Point-Contact-Andreev-Reflection~\cite{Soulen1998} (PCAR) on pure YBCO, or $\mathbf{(X,Y)=(0,0)}$, and $\mathbf{(X,Y)=(0.32,0.32)}$.}
A ZBCP in PCAR is a signature of a D-wave phase. No ZBCP is expected for an S-wave phase. This experiment searches for a sign change in the superconducting gap. (a) A ZBCP is seen for pure YBCO, as expected, since it has a D-wave gap. (b) No ZBCP is seen for $(X,Y)=(0.32,0.32)$, suggesting it has an S-wave gap.
These curves are representative of several measurements on each sample.}
\label{pcar}
\end{figure}

The second experiment looked for a sign change in the superconducting gap using Point-Contact-Andreev-Reflection~\cite{BTK1982} (PCAR).
PCAR measures the tunneling current from a normal metal (in this case Cu) point contact into the superconductor. The normal metal has a continuum of states in the neighborhood of the Fermi level, whereas the superconductor has its energy gap centered on its Fermi level. Thus no normal current can tunnel from the normal level into the superconductor for bias voltages less than half the gap. However, there are normal states in a D-wave superconductor in the regions where its gap changes sign, and electrons from the normal metal tip can tunnel into these states, thereby showing a sharp peak near zero bias voltage, known as the zero-bias conductance peak (ZBCP)~\cite{Tanaka1995,Kashiwaya1996}. An S-wave gap has no sign change and hence no ZBCP in PCAR. See Figure~\ref{pcar}.

The figures shows representative conductance plots from many spectra for $\mathrm{(0,0)}$ and $\mathrm{(0.32,0.32)}$. Since $\mathrm{(0,0)}$, pure YBCO, is D-wave, a ZBCP is seen, as expected. For heavily counter-doped $\mathrm{(0.32,0.32)}$, no ZBCP was found. We conclude that PCAR suggests highly counter-doped YBCO is an S-wave superconductor.

\subsection{Evolution of the low-temperature penetration depth}

\begin{figure}[!ht]
\centering \includegraphics[width=130mm]{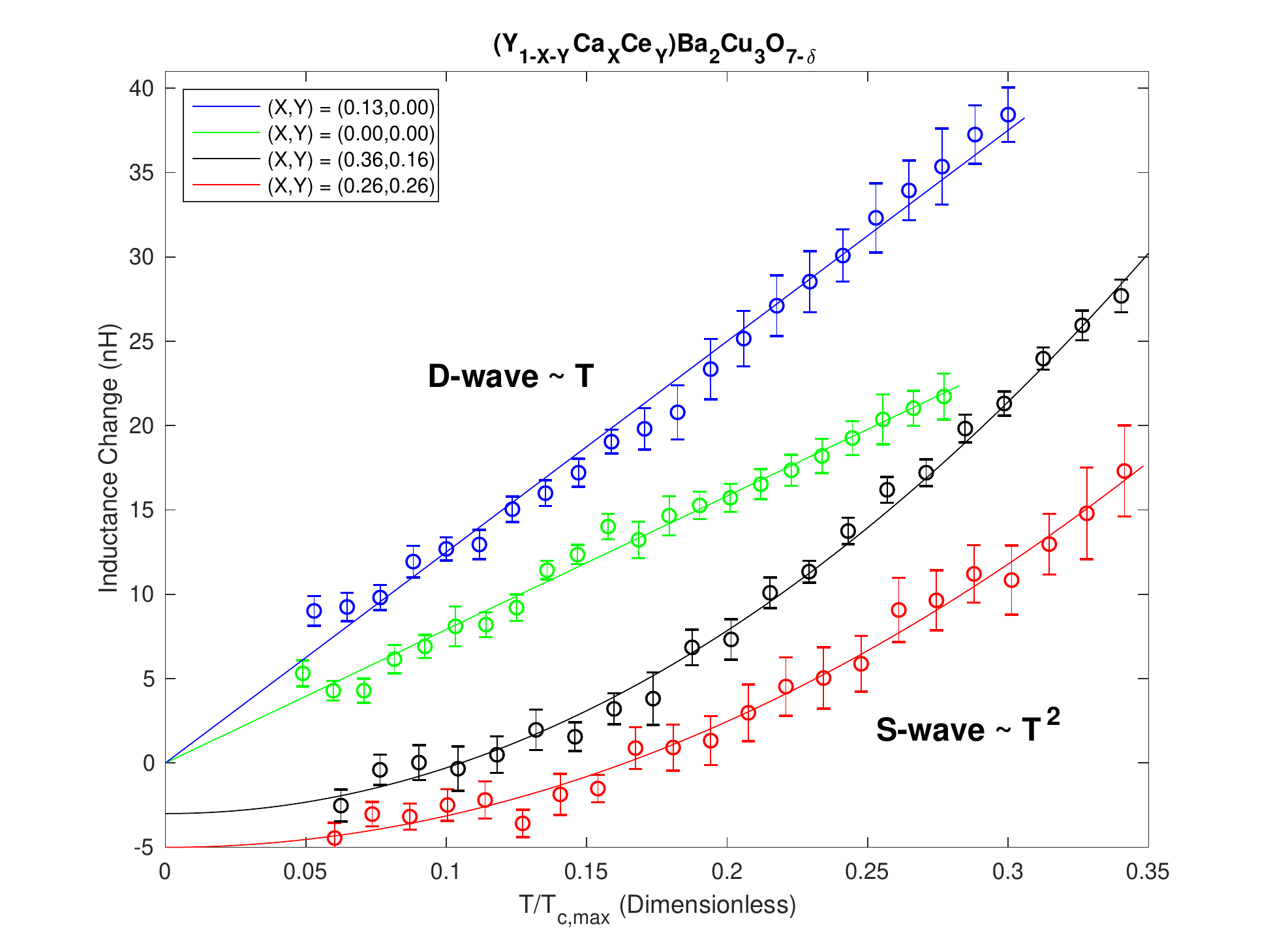}
\caption{
\textbf{Evolution of the measured superconducting penetration depth change as a function of the normalized temperature, as seen by the change in inductance.}
The figure shows the change in inductance for two known D-wave gap phases, pure YBCO, $\mathrm{(X,Y)=(0,0)}$ (green), $\mathrm{(X,Y)=(0.13,0.0)}$ (blue), and two phases with $\mathrm{(X,Y)=(0.36,0.16)}$ (black) and $\mathrm{(0.26,0.26)}$ (red). For clarity, the black data are shifted downward by $3\ \mathrm{nH}$ and the red data are shifted downward by $5\ \mathrm{nH}$.
The data points are the measured inductance with $\pm 3\sigma$ error bars. The x-axis is the ratio $\mathrm{T/T_{c,max}}$ where $\mathrm{T_{c,max}}$ is the maximum superconducting temperature from Figure~\ref{tcplot}. For all curves, the change in inductance was measured from $\mathrm{4\ K}$ to $\mathrm{26\ K}$.
In the Supplement~\cite{Supplement}, we show that the measured change in inductance is proportional to the change in penetration depth of the superconducting sample, and also estimate many sources of errors in this measurement. The extrapolated $T=0\ \mathrm{K}$ inductance values are
$19.276\ \mathrm{\mu H}$,
$20.623\ \mathrm{\mu H}$,
$20.349\ \mathrm{\mu H}$, and
$21.243\ \mathrm{\mu H}$
for $(0,0)$, $(0.13,0)$, $(0.36,0.16)$, and $(0.26,0.26)$, respectively.
The solid blue and green curves are linear in $T$ fits and the black and red curves are $T^2$ fits to the data. A linear $T$ evolution of the penetration depth is expected for a D-wave gap and $T^2$ is expected for an S-wave gap.
The green and blue curves show that YBCO with $\mathrm{(X,Y)=(0.0,0.0)}$ and $\mathrm{(0.13,0.0)}$ are D-wave, as expected. The black and red curves suggest that $\mathrm{(X,Y)=(0.36,0.16)}$ and $\mathrm{(0.26,0.26)}$ are S-wave gap phases of YBCO.
}
\label{lambda}
\end{figure}

The third experiment searched for nodes in the superconducting gap by the measuring the evolution of the superconducting penetration depth, $\mathrm{\lambda}$, as a function of temperature, T. A D-wave gap has nodes leading to $\mathrm{\lambda\sim T}$.
An S-wave superconductor in the London limit (short coherence length) has $\mathrm{\lambda\sim T^2}$ because it does not have nodes~\cite{Tinkham2004}.

Figure~\ref{lambda} shows the changes in inductance, $L$, of a pancake coil placed on top of $\mathrm{(X,Y)}$ samples for $\mathrm{(X,Y)=(0,0)}$, $\mathrm{(0.13,0)}$, $\mathrm{(0.36,0.16)}$, and $\mathrm{(0.32,0.32)}$. Changes in $L$ are proportional to changes in the superconducting penetration depth, $\lambda$~\cite{Supplement}.  The $\mathrm{(0,0)}$ and $\mathrm{(0.13,0)}$ are known D-wave gap materials. Hence, we expect that their $L$ changes linearly with $T$, as observed. We find $L\sim T^2$ for $\mathrm{(0.36,0.16)}$ and $\mathrm{(0.26,0.26)}$ leading to $\lambda\sim T^2$ for both samples, as expected for an S-wave gap superconductor. 

The changes in $L$ in Figure~\ref{lambda} are several 10s of nanoHenrys. The magnitude of the extrapolated $L$ at $T=0$ is $\approx 20\ \mathrm{\mu H}$ for the four curves. A $1\ \mathrm{nH}$ change is a relative $L$ change of $\sim 5\times 10^{-5}$, making this experiment the most difficult of the three experiments in Figures~\ref{tcplot}, \ref{pcar}, and \ref{lambda}. A detailed description of this experiment and estimates of many potential errors is in the Supplement~\cite{Supplement}.

A possible D-wave gap explanation for $\lambda\sim T^2$ exists. Hirschfeld et al.~\cite{Hirschfeld1993} showed that a D-wave superconductor with non-magnetic impurities can lead to $\lambda\sim T^2$ for $T<T^*$, where $T^*$ depends on the magnitude of single impurity scattering (in our case, a single Ca or Ce atom) and the ratio of $\mathrm{T_{c,max}}$ for (X,Y) to $\mathrm{T_{c,max}}$ for $\mathrm{(0,0)}$ (pure YBCO). For $T>T^*$, the theory predicts $\lambda\sim T$. Since Ca and Ce impurities reside at the Y site in YBCO and this site is not in the $\mathrm{CuO_2}$ planes, where most of the density of the metallic band is located, the magnitude of single impurity scattering is small~\cite{Born2009}. Extended Data Figure~\ref{tstar} shows that the theory prediction~\cite{Hirschfeld1993} for weak (Born) scattering plus the $\mathrm{T_{c,max}}$ values measured in Figure~\ref{tcplot} lead to the conclusion that $T^*\ll 1\ \mathrm{K}$ for our experiment. Hence, a D-wave superconductor with impurity scattering does not explain the observed $\lambda\sim T^2$ up to $\mathrm{26\ K}$ as seen for $\mathrm{(0.36,0.16)}$ and $\mathrm{(0.26,0.26)}$.

\subsection{Summary of the three experiments}

Table~\ref{list} summarizes the findings from the three experiments in Figures~\ref{tcplot}, \ref{pcar}, and \ref{lambda}. In all three experiments, the results favored a crossover from a D-wave gap at low counter-doping to an S-wave gap at high counter-doping. These results imply that Figure~\ref{tcdvs}b is correct---a technologically useful S-wave superconducting gap YBCO phase resides at $\mathrm{\sim 70\ K}$ in YBCO heavily counter-doped with Ca and Ce impurities.

We found two papers in the literature where a crossover from D-wave superconductivity to S-wave was seen. First, in 2001, Yeh et al.~\cite{Yeh2001} observed a $d+s$ superconducting gap symmetry for highly doped
$\mathrm{(Y_{0.7}Ca_{0.3})Ba_2Cu_3O_{7-\delta}}$ with $\mathrm{T_c=78\pm 2\ K}$. In our notation, this sample has $\mathrm{(X,Y)=(0.3,0.0)}$. From Figure~\ref{tcplot}, the $\mathrm{T_c}$ measured by Yeh et al., falls on the blue line for $\mathrm{T_{c,max}}$ and is in the crossover region between a D-wave gap to an S-wave gap.

Second, in 2012, Reid et al.~\cite{Taillefer2012} found a crossover from a D-wave superconducting gap in the iron-pnictide $\mathrm{KFe_2As_2}$ to an S-wave gap in $\mathrm{(Ba_{0.6}K_{0.4})Fe_2As_2}$.
The authors of this paper attribute the gap symmetry change to a change in the Fermi surface and Fermi level between the two pnictide samples. 
In our samples, we believe the gap symmetry has changed without altering the Fermi surface or the Fermi level.

\FloatBarrier

\begin{table}[h]
\caption{
\textbf{Summary of the  results of thre experiments on the gap symmetry.}
All three experiments suggest that Ca and Ce doped YBCO has changed from D-wave to an S-wave phase at high Ca and Ce counter-doping.}
\label{list}
\begin{tabular}{c|c|c|c|c|c}
 Figure & Physical & Experiment & Expected for & Expected for & Result \\
        & Property &            & D-wave       & S-wave       &        \\
\hline
\hline
   \ref{tcplot}    & Superconducting & $T_{c}$ change with & $T_{c}$ decreases with & $T_{c}$ weakly dependent & S-wave \\
        & gap magnitude   & non-magnetic counter-doping & increasing counter-doping & on counter-doping & \\
\hline
  \ref{pcar}     & Superconducting & Point-Contact- & Zero-Bias Conductance & No ZBCP & S-wave \\
        & gap phase  & Andreev-Reflection (PCAR) & Peak (ZBCP) &  & \\
\hline
  \ref{lambda}     & Superconducting & Penetration depth, $\lambda$, & $\lambda\sim T$ & $\lambda\sim T^2$ & S-wave \\
        & zero-energy excitations & dependence at & & & \\
        &                         & low temperature, $T$ & & & \\
\hline
\end{tabular}
\end{table}

\FloatBarrier

\section{Conclusions}

The intent of this paper was to search for an S-wave gap symmetry YBCO phase beneath the known D-wave gap YBCO phase at $\mathrm{\sim 90\ K}$ and determine its superconducting transition temperature. Our conjecture was that an S-wave YBCO phase was at $\mathrm{\sim 70\ K}$, and if true, will have huge implications for making high-field magnets using poly-crystalline superconducting wires. To test this conjecture, we counter-doped YBCO with Ca and Ce impurities, performed three experiments to study the superconducting gap symmetry, and found evidence suggesting that an S-wave gap symmetry phase does exist at $\mathrm{\sim 70\ K}$ in all three experiments (see Table~\ref{list}).

A potential S-wave gap symmetry crossover in Ca and Ce counter-doped cuprate, $\mathrm{YBa_{2}Cu_{4}O_{8}}$ (Y124) should be studied. Y124 is a stoichiometric crystal that is much more three-dimensional than YBCO (vastly improved conduction normal to its $\mathrm{CuO_2}$ plane compared to YBCO). Y124 is intrinsically underdoped. Doping with 0.1 Ca brings Y124 up to optimal doping (highest $\mathrm{T_c}$). Hence, counter-doping with 0.1 more Ca than Ce is desired for the highest $\mathrm{T_c}$.

Poly-crystalline counter-doped wires of Y124 will be mechanically strong. Using the metallic precursor method~\cite{Masur1994} to form poly-crystalline wires of Y124 is already known to lead to grain alignments $\mathrm{<10^o}$ normal to the $\mathrm{CuO_2}$ planes and grain alignments $\mathrm{<15^o}$ in the planes. While these grain mis-alignments made D-wave poly-crystalline Y124 impractical, a counter-doped Y124 that becomes S-wave may have a very large supercurrent density using this mature manufacturing process.

The results in this paper were obtained on poly-crystalline counter-doped samples. Ideally, one would like to repeat these experiments and additional experiments on single-crystals. The change in critical current density as a function of grain misalignment should also be measured to determine how much supercurrent can be transported in poly-crystalline wires.

Poly-crystalline wires should be synthesized and characterized. Two methods exist for making poly-crystalline wires from cuprates: Powder-in-Tube~\cite{Hellstrom2019,Paturi2004,Chaffron1991} and metallic precursors~\cite{Masur1994,Masur1993a,Masur1993b,Masur1995}. If there is a large improvement in the supercurrent carrying ability of these wires, then an enormous opportunity exists for creating new and useful practical wires for high-magnetic field applications on a short timescale.

\bibliography{gap.bbl}

\begin{thebibliography}{10}
\expandafter\ifx\csname url\endcsname\relax
  \def\url#1{\texttt{#1}}\fi
\expandafter\ifx\csname urlprefix\endcsname\relax\def\urlprefix{URL }\fi
\providecommand{\bibinfo}[2]{#2}
\providecommand{\eprint}[2][]{\url{#2}}

\bibitem[Larbelestier, D. and Gurevich, A. and Feldmann, D.M. and Polyanskii, A.]{Larbelestier2001}
Larbelestier, D., et~al.
\newblock {\em High-Tc superconducting materials for electric power applications}.
\newblock \href{https://www.nature.com/articles/35104654}
{Nature {\bf 414}, 368--377 (2001)}

\bibitem[Hirschfeld, Peter J. and Goldenfeld, Nigel]{Hirschfeld1993}
Hirschfeld, P. and Goldenfeld, N.
\newblock {\em Effect of strong scattering on the low-temperature penetration depth of a d-wave superconductor}.
\newblock \href{https://link.aps.org/doi/10.1103/PhysRevB.48.4219}
{Physical Review B {\bf 48}, 4219--4222 (1993)}.

\bibitem{Abrikosov1961}
Abrikosov, A.~A. and Gor'kov, L.~P.
\newblock {\em Contribution to the Theory of Superconducting Alloys with Paramagnetic Impurities}.
\newblock \href{https://www.osti.gov/biblio/4097498}
{Soviet Physics--JETP {\bf 12}, 1243--1253 (1961)}.

\bibitem{Anderson1959}
Anderson, P.~W.
\newblock {\em Theory of dirty superconductors}
\newblock \href{https://doi.org/10.1016/0022-3697(59)90036-8}
{Journal of Physics and Chemistry of Solids {\bf 11}, 26--30 (1959)}

\bibitem[J. Tahir-Kheli (2017)]{JTK2017}
Tahir-Kheli, J.
\newblock {\em Latent Room-Temperature T$_{\rm c}$ in Cuprate Superconductors}.
\newblock \href{https://doi.org/10.48550/arXiv.1702.05001}
{arxiv:1702.05001 (2017)}.

\bibitem{Tinkham2004}
Tinkham, M.
\newblock {\em Introduction to Superconductivity: Second Edition}.
{Dover Publications, page 104, (2004)}.

\bibitem{Williams1998}
Williams, G.~V.~M. et~al.
\newblock {\em NMR studies of overdoped ${\mathrm{Y}}_{1\ensuremath{-}x}{\mathrm{Ca}}_{x}{\mathrm{Ba}}_{2}{\mathrm{Cu}}_{3}{\mathrm{O}}_{7\mathrm{\ensuremath{-}}\mathrm{\ensuremath{\delta}}}$}.
\newblock \href{https://link.aps.org/doi/10.1103/PhysRevB.57.8696}
{Physical Review B {\bf 57}, 8696--8701 (1998)}.

\bibitem{Naqib2003}
Naqib, S.~H. et~al.
\newblock {\em Temperature dependence of electrical resistivity of high-t-c cuprates - from pseudogap to overdoped regions}.
\newblock \href{http://dx.doi.org/10.1016/S0921-4534(02)02330-4}
{Physica C-Superconductivity and Its Applications {\bf 387}, 365--372 (2003)}.

\bibitem{Jin1991}
Jin, S. et~al.
\newblock {\em Superconducting properties of ${\mathrm{YBa}}_{2}{\mathrm Cu}_{3}{\mathrm O}_{7{\mathrm -\delta}}$ with partial rare earth substitution}.
\newblock \href{https://doi.org/10.1016/0921-4534(91)90795-Z}
{Physica C: Superconductivity {\bf 173}, 75--79 (1991)}.

\bibitem[Tallon, J. L. and Bernhard, C. and Shaked, H. and Hitterman, R. L. and Jorgensen, J. D.]{Tallon1995}
Tallon, J.~L. et~al.
\newblock {\em Generic superconducting phase behavior in high-${\mathit{T}}_{\mathit{c}}$ cuprates: ${\mathit{T}}_{\mathit{c}}$ variation with hole concentration in ${\mathrm{YBa}}_{2}$${\mathrm{Cu}}_{3}$${\mathrm{O}}_{7\mathrm{\ensuremath{-}}\mathrm{\ensuremath{\delta}}}$}.
\newblock \href{https://link.aps.org/doi/10.1103/PhysRevB.51.12911}
{Physical Review B {\bf 51}, 12911--12914 (1995)}.

\bibitem[Sup()]{Supplement}
{\em Supplemental Material}.
\newblock See Supplemental Material for further details.

\bibitem[Presland, Tallon, Buckly, Liu, Flower]{Presland1991}
Presland, M.~R. et~al.
\newblock {\em General trends in oxygen stoichiometry effects on $\mathit{T_c}$ in Bi and Tl superconductors}.
\newblock \href{https://doi.org/10.1016/0921-4534(91)90700-9}
{Physica C: Superconductivity {\bf 176}, 95--105 (1991)}.

\bibitem[Obertelli et~al (1992)]{Obertelli1992}
Obertelli, S.~D. et~al.
\newblock {\em Systematics in the thermoelectric power of high-${\mathit{T}}_{\mathit{c}}$ oxides}.
\newblock \href{https://link.aps.org/doi/10.1103/PhysRevB.46.14928}
{Physical Revew B {\bf 46}, 14928--14931 (1992)}.

\bibitem{BTK1982}
Blonder, G.~E. et~al.
\newblock {\em Transition from metallic to tunneling regimes in superconducting microconstrictions: Excess current, charge imbalance, and supercurrent conversion}.
\newblock \href{https://link.aps.org/doi/10.1103/PhysRevB.25.4515}
{Physical Review B {\bf 25}, 4515--4532 (1982)}.

\bibitem{Tanaka1995}
Tanaka, Y. and Kashiwaya, S.
\newblock {\em Theory of Tunneling Spectroscopy of $\mathit{d}$-Wave Superconductors}.
\newblock \href{https://link.aps.org/doi/10.1103/PhysRevLett.74.3451}
{Physical Review Letters {\bf 74}, 3451--3454 (1995)}.

\bibitem{Kashiwaya1996}
Kashiwaya, S. et~al.
\newblock {\em Theory for tunneling spectroscopy of anisotropic superconductors}.
\newblock \href{https://link.aps.org/doi/10.1103/PhysRevB.53.2667}
{Physical Review B {\bf 53}, 2667--2676 (1996)}.

\bibitem[Alloul, H. and Bobroff, J. and Gabay, M. and Hirschfeld, P. J.]{Born2009}
Alloul, H. et~al.
\newblock {\em Defects in correlated metals and superconductors}.
\newblock \href{https://link.aps.org/doi/10.1103/RevModPhys.81.45}
{Reviews of Modern Physics {\bf 81}, 45--108 (2009)}.

\bibitem{Yeh2001}
Yeh, N.~C. et~al.
\newblock {\em Evidence of Doping-Dependent Pairing Symmetry in Cuprate Superconductors}.
\newblock \href{https://doi.org/10.1103/PhysRevLett.87.087003}
{Physical Review Letters {\bf 87} 087003 (4pp) (2001)}.

\bibitem{Taillefer2012}
Reid, J-Ph. et~al.
\newblock {\em From d-wave to s-wave pairing in the iron-pnictide superconductor (Ba,K)Fe$_2$As$_2$}.
\newblock \href{https://doi.org/10.1088/0953-2048/25/8/084013}
{Superconductor Science and Technology {\bf 25}, 084013 (10pp) (2012)}.

\bibitem[R.J. Soulen, Jr., J.M. Byers, M.S. Osofsky, B. Nadgorny, T. Ambrose, S.F. Cheng, P.R. Broussard, C.T. Tanaka, J. Nowak, J. S. Moodera, A. Barry, and J.M.D. Coey]{Soulen1998}
Soulen, R.~J. et~al.
\newblock {\em Measuring the Spin Polarization of a Metal with a Superconducting Point Contact}.
\newblock \href{https://doi.org/10.1126/science.282.5386.85}
{Science {\bf 282}, 85--88 (1998)}.

\bibitem{Masur1994}
Masur, L.J. et~al.
\newblock {\em Bi-axial texture in $Ca_{0.1}Y_{0.9}Ba_2Cu_4O_8$ composite wires made by metallic precursors}.
\newblock \href{https://doi.org/10.1016/0921-4534(94)90840-0}
{Physica C: Superconductivity {\bf 230}, 274--282 (1994)}.

\bibitem{Hellstrom2019}
Zhang, Z. et~al.
\newblock {\em Investigation of the melt-growth process of
$YbBa_{2}Cu_{3}O_{7-\delta}$ powder in Ag-sheathed tapes}.
\newblock \href{https://doi.org/10.1039/C8CE02079E}
{CrystEngComm {\bf 21}, 1369--1377 (2019)}.

\bibitem{Paturi2004}
Paturi, P. et~al.
\newblock {\em Texture of YBCO/Ag PIT-tapes}.
\newblock \href{https://doi.org/10.1016/j.physc.2004.03.170}
{Physica C: Superconductivity {\bf 408-410}, 935--936}.

\bibitem{Chaffron1991}
Chaffron, L. and R\'{e}gnier, P.
\newblock {\em Present state of our research on silver sheathed REBa2Cu3O7 superconductors}.
\newblock \href{https://doi.org/10.1088/0953-2048/4/1S/068}
{Superconductor Science and Technology {\bf 4}, S247--S249 (1991)}.

\bibitem{Masur1993a}
Otto, A. et~al.
\newblock {\em Properties of high-$T_c$ wires made by the metallic precursor process}.
\newblock \href{https://doi.org/10.1007/BF03222435}
{JOM {\bf 45}, 48--52 (1993)}.

\bibitem{Masur1993b}
Otto, A. et~al.
\newblock {\em Multifilamentary Bi-2223 composite tapes made by a metallic precursor route}.
\newblock \href{https://doi.org/10.1109/77.233845}
{IEEE Transactions on Applied Superconductivity {\bf 3} 915--922 (1993)}.

\bibitem{Masur1995}
Otto, A. et~al.
\newblock {\em Progress towards a long length metallic precursor process for multifilament Bi-2223 composite superconductors}.
\newblock \href{https://doi.org/10.1109/77.402766}
{IEEE Transactions on Applied Superconductivity {\bf 5} 1154--1157 (1995)}.

\bibitem[Hardy, W. N. and Bonn, D. A. and Morgan, D. C. and Liang, Ruixing and Zhang, Kuan]{Hardy1993}
Hardy, W.~N., et~al.
\newblock {\em Precision measurements of the temperature dependence of \ensuremath{\lambda} in ${\mathrm{YBa}}_{2}$${\mathrm{Cu}}_{3}$${\mathrm{O}}_{6.95}$: Strong evidence for nodes in the gap function}.
\newblock \href{https://link.aps.org/doi/10.1103/PhysRevLett.70.3999}
{Physical Review Letters {\bf 70}, 3999--4002 (1993)}.

\end{thebibliography}


\begin{thebibliography}{10}
\expandafter\ifx\csname url\endcsname\relax
  \def\url#1{\texttt{#1}}\fi
\expandafter\ifx\csname urlprefix\endcsname\relax\def\urlprefix{URL }\fi
\providecommand{\bibinfo}[2]{#2}
\providecommand{\eprint}[2][]{\url{#2}}

\bibitem[Hardy, W. N. and Bonn, D. A. and Morgan, D. C. and Liang, Ruixing and Zhang, Kuan]{Hardy1993s}
Hardy, W.~N., et~al.
\newblock {\em Precision measurements of the temperature dependence of \ensuremath{\lambda} in ${\mathrm{YBa}}_{2}$${\mathrm{Cu}}_{3}$${\mathrm{O}}_{6.95}$: Strong evidence for nodes in the gap function}.
\newblock \href{https://link.aps.org/doi/10.1103/PhysRevLett.70.3999}
{Physical Review Letters {\bf 70}, 3999--4002 (1993)}.

\bibitem[King (1921)]{King1921}
King, L.~V.
\newblock {\em New formulae for the numerical calculation of the mutual induction of coaxial circles}.
\newblock \href{https://www.jstor.org/stable/93861}
{Proceedings of the Royal Society of London. Series A, Containing Papers of a Mathematical and Physical Character, {\bf 100}, 60--66 (1921)}

\bibitem[Obertelli et~al (1992)]{Obertelli1992}
Obertelli, S.~D. et~al.
\newblock {\em Systematics in the thermoelectric power of high-${\mathit{T}}_{\mathit{c}}$ oxides}.
\newblock \href{https://link.aps.org/doi/10.1103/PhysRevB.46.14928}
{Physical Revew B {\bf 46}, 14928--14931 (1992)}.

\bibitem[Presland, Tallon, Buckly, Liu, Flower]{Presland1991}
Presland, M.~R. et~al.
\newblock {\em General trends in oxygen stoichiometry effects on $\mathit{T_c}$ in Bi and Tl superconductors}.
\newblock \href{https://doi.org/10.1016/0921-4534(91)90700-9}
{Physica C: Superconductivity {\bf 176}, 95--105 (1991)}.

\bibitem{Abrikosov1961}
Abrikosov, A.~A. and Gor'kov, L.~P.
\newblock {\em Contribution to the Theory of Superconducting Alloys with Paramagnetic Impurities}.
\newblock \href{https://www.osti.gov/biblio/4097498}
{Soviet Physics--JETP {\bf 12}, 1243--1253 (1961)}.

\bibitem[Hirschfeld, Peter J. and Goldenfeld, Nigel]{Hirschfeld1993}
Hirschfeld, P. and Goldenfeld, N.
\newblock {\em Effect of strong scattering on the low-temperature penetration depth of a d-wave superconductor}.
\newblock \href{https://link.aps.org/doi/10.1103/PhysRevB.48.4219}
{Physical Review B {\bf 48}, 4219--4222 (1993)}.

\bibitem{Anderson1959}
Anderson, P.~W.
\newblock {\em Theory of dirty superconductors}
\newblock \href{https://doi.org/10.1016/0022-3697(59)90036-8}
{Journal of Physics and Chemistry of Solids {\bf 11}, 26--30 (1959)}

\bibitem[Clark, A.~F. and Childs, G.~E. and Wallace, G.~H]{Clark1970}
Clark, A.~F. et~al.
\newblock{\em Electrical resistivity of some engineering alloys at low temperatures}.
\newblock \href{https://doi.org/10.1016/0011-2275(70)90056-1}
{Cryogenics, {\bf 10} 295--305 (1970)}

\bibitem[Tuttle, J. and Canavan, E. and DiPirro, M.]{Tuttle2010}
Tuttle, J., et~al.
\newblock{\em Thermal and electrical conductivity measurements of CDA 510 phosphor-bronze}.
\newblock \href{https://doi.org/10.1063/1.3402333}
{AIP Conference Proceedings, {\bf 55} 55--62 (2010)}

\bibitem[Simon, N.J. and Drexler, E.S. and Reed, R.P.]{NIST-Cu1992}
Simon, N.~J., et~al.
\newblock {\em Properties of Cu and Cu alloys at cryogenic temperatures}.
\newblock \href{https://doi.org/10.2172/5340308}
{NIST Monograph {\bf 177}, Chapter 20, page 21 (1992). U.S. Government Printing Office, Washington, DC 20402-9325}

\bibitem[Larbelestier, D. and Gurevich, A. and Feldmann, D.M. and Polyanskii, A.]{Larbelestier2001}
Larbelestier, D., et~al.
\newblock {\em High-Tc superconducting materials for electric power applications}.
\newblock \href{https://www.nature.com/articles/35104654}
{Nature {\bf 414}, 368--377 (2001)}

%
%

\end{thebibliography}

\setcounter{figure}{0}
\captionsetup[figure]{labelfont={bf},name={Ext. Data Fig.},labelsep=bar,justification=raggedright,font=small}

\section*{Extended Data Figures}
\FloatBarrier

\begin{figure}[!ht]
\centering \includegraphics[width=130mm]{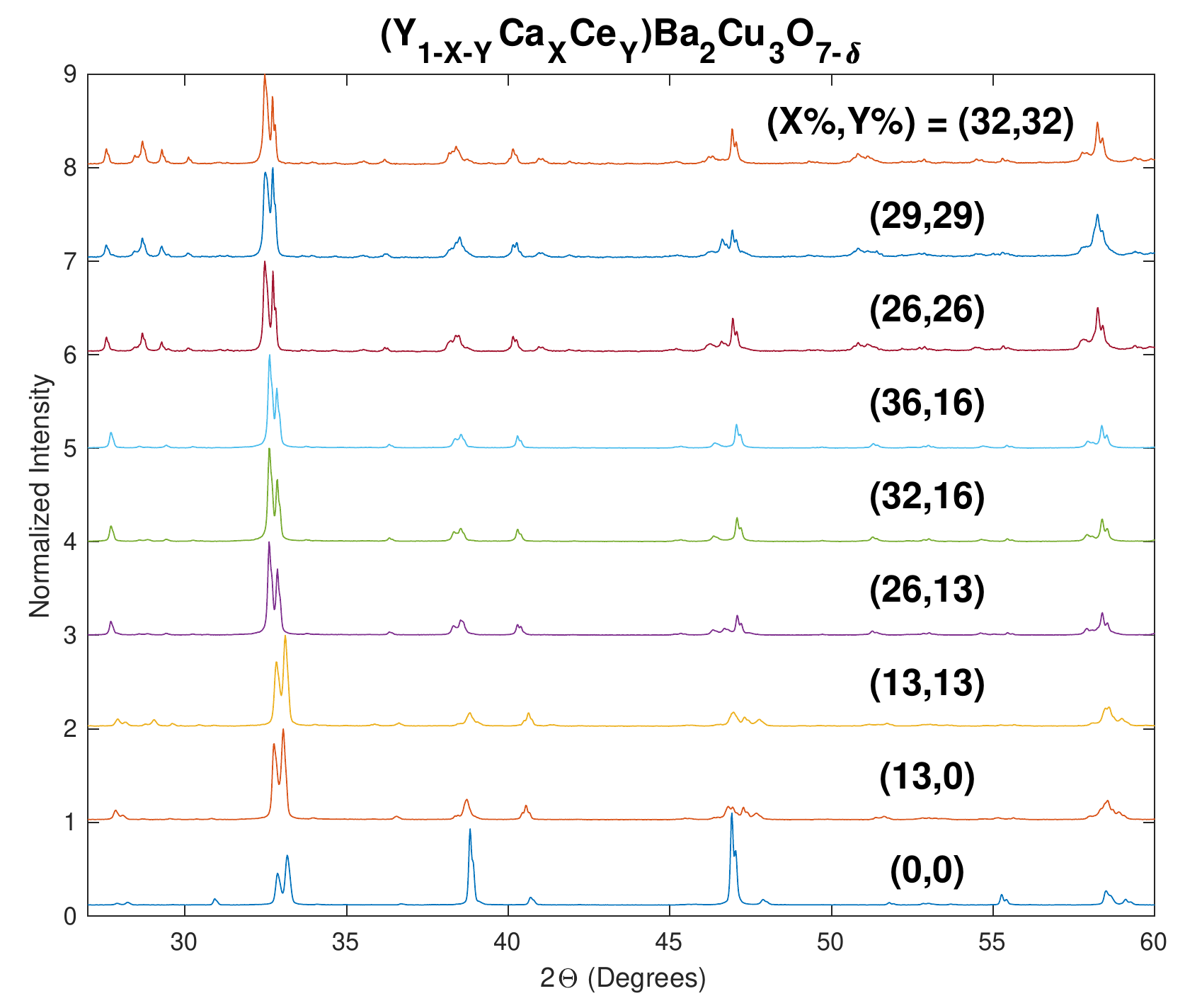}
\caption{
\textbf{X-ray diffraction (XRD) patterns for
$\mathbf{(Y_{1-x-y}Ca_{x}Ce_{y})Ba_{2}Cu_{3}O_{7-\delta}}$
\ [hereinafter, denoted (X,Y)] for (X,Y) ranging from pure YBCO, or (X,Y) = (0,0), up to (X,Y)=(0.32, 0.32).}
There is a small amount of alien phase seen below $\mathrm{30^o}$ for the three $\mathrm{X=Y}$ samples, $\mathrm{(0.26,0.26)}$, $\mathrm{(0.29,0.29)}$, and $\mathrm{(0.32,0.32)}$. Otherwise, the samples are single-phase. See Extended Data Figures~\ref{bsesem2626} and \ref{edx2626} and Extended Data Tables~\ref{rietveld} and \ref{phasecomp}.
}
\label{xrd}
\end{figure}

\begin{table}[tbp]
\caption{
\textbf{Rietveld x-ray occupation of Ca and Ce atoms on Y for (X,Y) = (0.26, 0.13), (0.32, 0.16), and (0.36, 0.16).}
These results show that the Ca and Ce atoms have substituted onto the Y site and that the measured composition is close to the desired composition.
}
\label{rietveld}
\begin{tabular}{c|cc|cc|cc|}
  & \multicolumn{2}{c|}{(0.26, 0.13)} & \multicolumn{2}{c|}{(0.32, 0.16)} & \multicolumn{2}{c|}{(0.36, 0.16)} \\
\hline
  Element & Expected & Refined & Expected & Refined & Expected & Refined \\
\hline
  Y       &  0.61    & 0.61    &  0.52    & 0.52     & 0.48   & 0.48     \\
  Ca      &  0.26    & 0.255(4)&  0.32    & 0.300(4) & 0.36   & 0.348(3) \\
  Ce      &  0.13    & 0.135(4)&  0.16    & 0.180(4) & 0.16   & 0.172(3) \\
  Ba      &  2       &  2      &  2       &  2       &  2     & 2        \\
  Cu      &  3       &  3      &  3       &  3       &  3     & 3        \\
  O       &  7       &  7      &  7       &  7       &  7     & 7        \\
\hline
\end{tabular}
\end{table}

\begin{table}[tbp]
\caption{
\textbf{Phase composition of (X,Y) = (0.26, 0.13), (0.32, 0.16), and (0.36, 0.16) in weight \% as determined by Rietveld x-ray diffraction.}
In these samples, the alien phase is $\mathrm{BaCuO_2}$.
}
\label{phasecomp}
\begin{tabular}{c|ccc|}
  & \multicolumn{3}{c|}{Phase Composition} \\
  & \multicolumn{3}{c|}{(weight \%)} \\
  & (0.26, 0.13) & (0.32, 0.16) & (0.36, 0.16) \\
\hline
 $\mathrm{(YCaCe)Ba_2Cu_3O_{7-\delta}}$ & 98.3(6) & 98.3(6) & 97.1(5) \\
 $\mathrm{BaCuO_2}$ & 1.66(6) & 1.67(6) & 2.85(5) \\
\hline
\end{tabular}
\end{table}

\begin{figure}[tbp]
\centering \includegraphics[width=150mm]{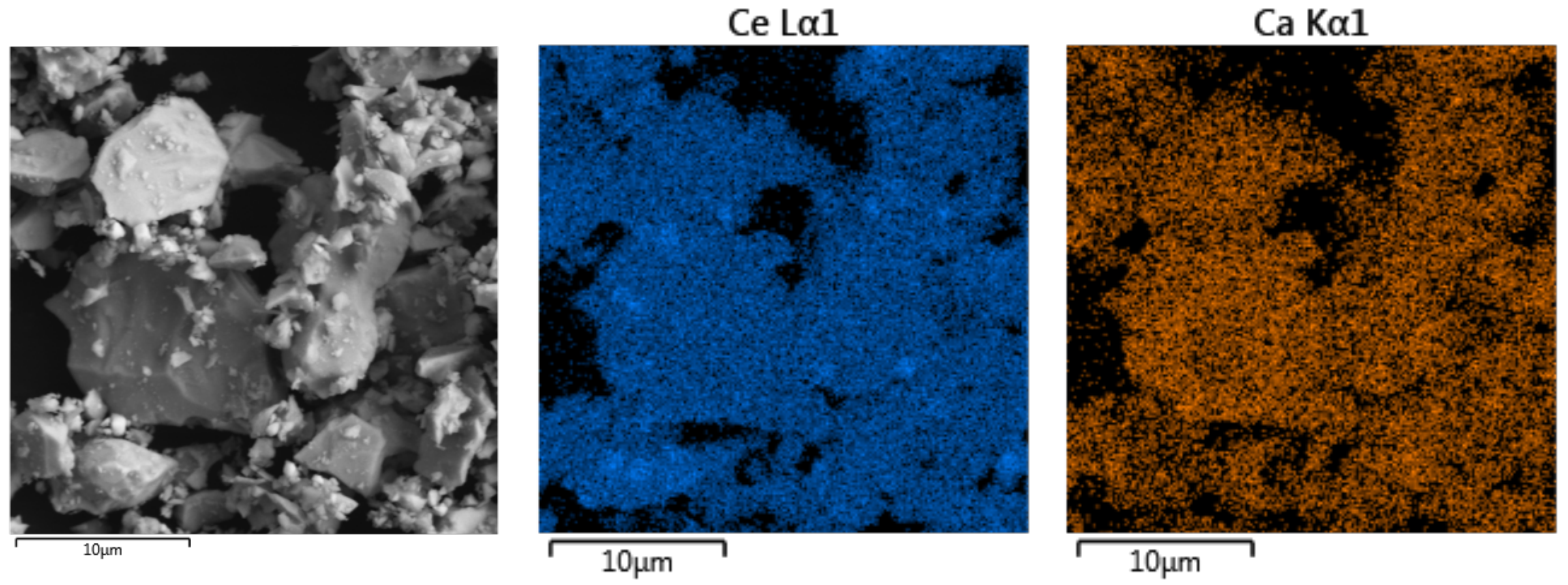}
\caption{
\textbf{Back-scattered scanning-electron-microscopy (BSE-SEM) and energy-dispersive X-ray (EDX) analysis of (X,Y) = (0.26, 0.26).
}
The BSE-SEM picture on the left shows no significant contrast differences that would suggest large grains of secondary phases. The Ce $\mathrm{L\alpha 1}$ and Ca $\mathrm{K\alpha 1}$ EDX distributions in the two right figures show that Ca and Ce are dispersed throughout the sample with similar distributions. This similarity implies that the grains have $\mathrm{X\approx Y}$. This finding is important in order to verify that the ``saturation" of the $\mathrm{T_c}$ values on the right-hand side of Figure~\ref{tcplot} are not due to an incorrect assignment of the $\mathrm{(X,Y)}$ values of the sample.
}
\label{bsesem2626}
\end{figure}

\begin{figure}[!h]
\centering \includegraphics[width=150mm]{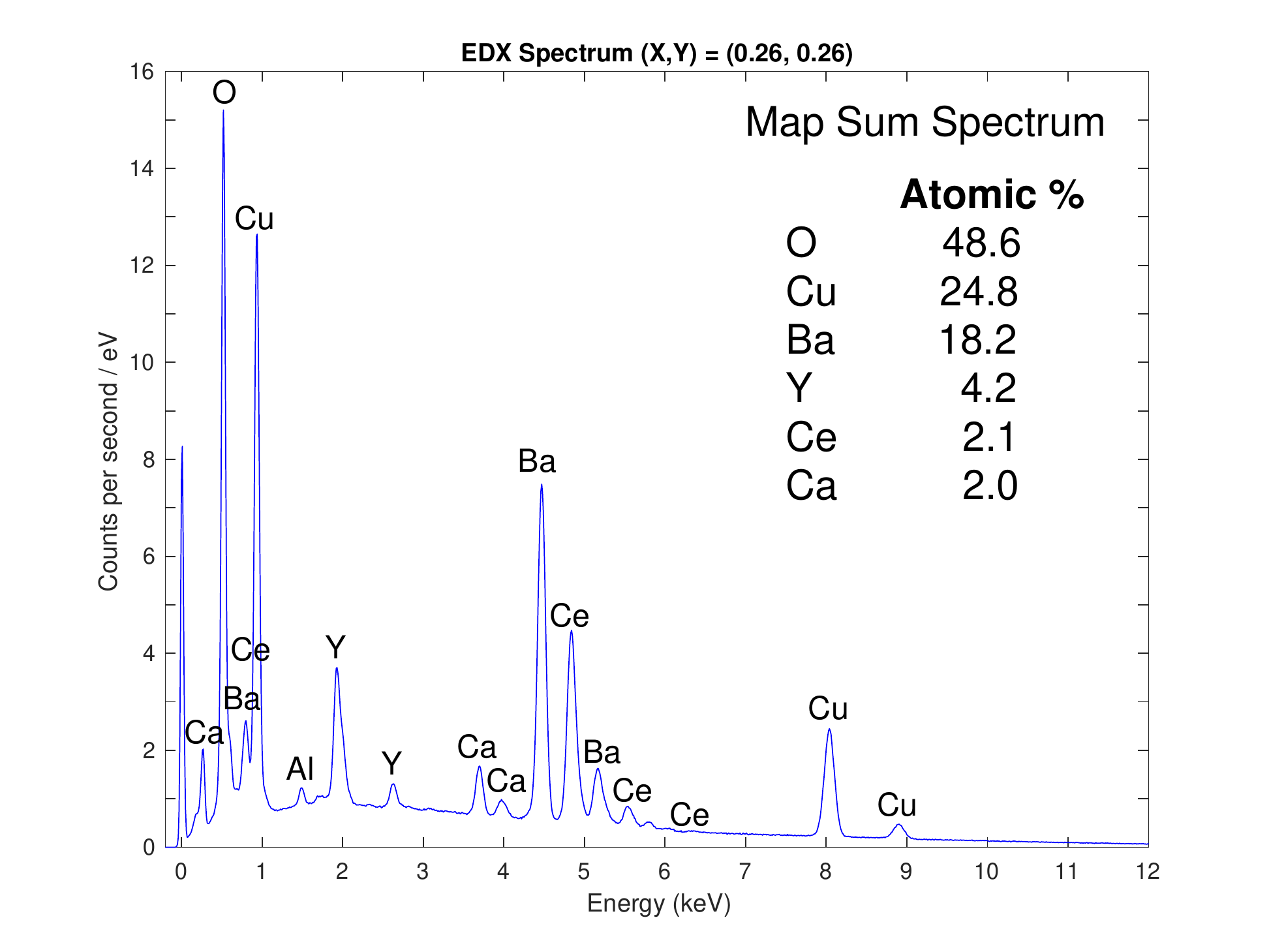}
\caption{
\textbf{EDX mass sum spectrum of an (X,Y) = (0.26, 0.26) sample.}
The atomic \% of each element is shown in the inset. From these values, we infer that $\mathrm{Ca:Ce=1:1}$, $\mathrm{Ca:Y=1:2}$, and $\mathrm{(Y,Ca,Ce):Ba:Cu:O=1:2:3:6}$ within experimental error. Hence, the synthesized $(X,Y)=(0.26,0.26)$ sample is close to the desired starting composition. This figure combined with Extended Figure~\ref{bsesem2626} shows that the ``saturation" of $T_c$ in Figure~\ref{tcplot} extends over a broad counter-doping range, as expected for an S-wave YBCO phase. The plot shows the spectrum up to $\mathrm{12\ keV}$. The data was taken up to $\mathrm{15\ keV}$. There were no large peaks above $\mathrm{12\ keV}$. The large peak at $\mathrm{0\ keV}$ is the instrumental response and does not correspond to any element. The Al peak at $\mathrm{\approx 1.5\ keV}$ is due to the sample holder.
}
\label{edx2626}
\end{figure}

\begin{figure}[tbp]
\centering \includegraphics[width=130mm]{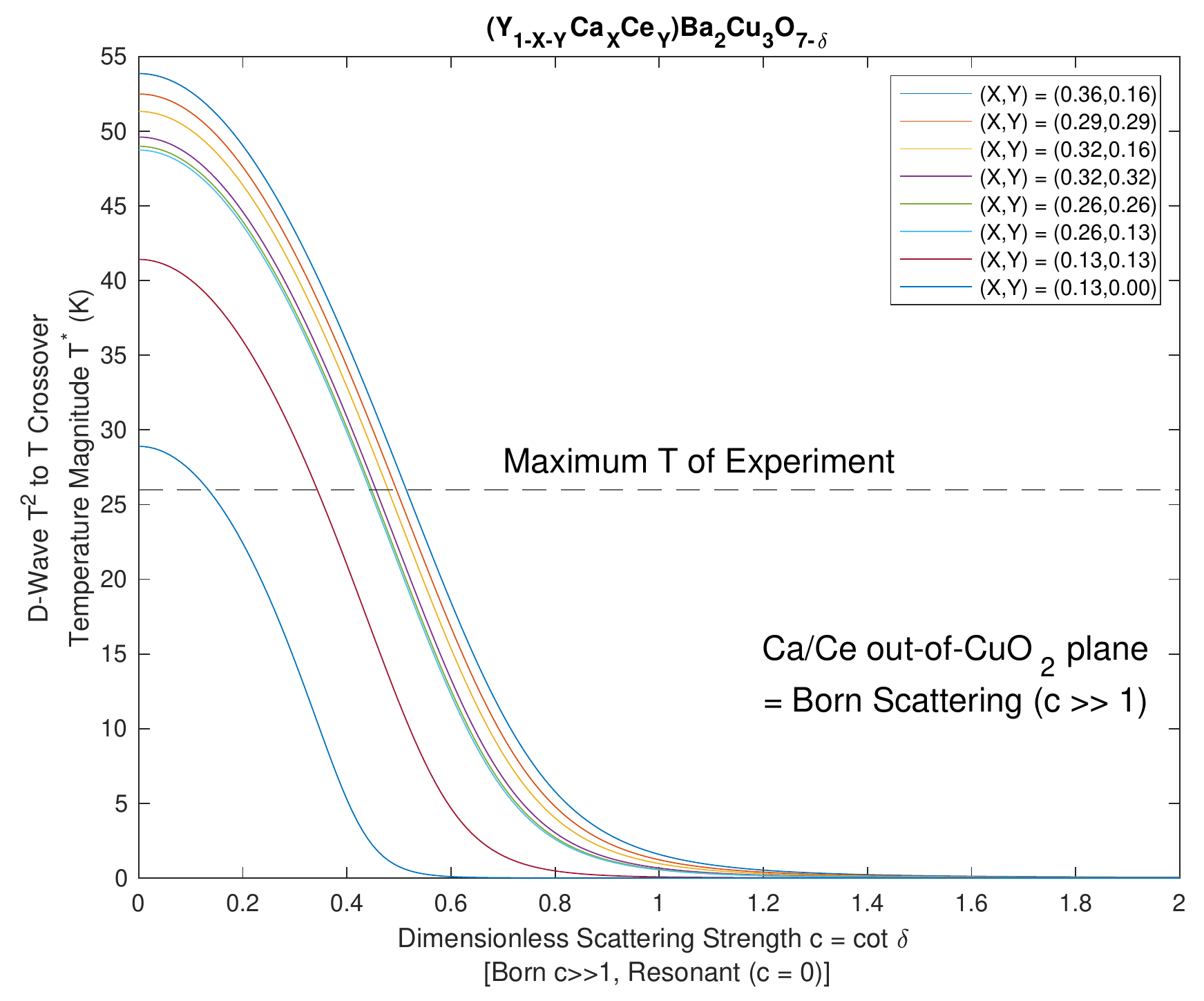}
\caption{
\textbf{Theoretical calculation of the temperature, $\mathbf{T^*}$, for the crossover from a $\mathbf{\lambda\sim T^2}$ penetration depth for $\mathbf{T<T^*}$ and a linear $\mathbf{\lambda\sim T}$ for $\mathbf{T>T^*}$ as a function of $\mathbf{c}$, the cotangent of the scattering phase shift, $\mathrm{\delta}$, per impurity (Ca or Ce) arising from non-magnetic counter-doping in a D-wave superconductor.
}
Non-magnetic impurity scattering in a D-wave superconductor leads to $\mathrm{\lambda\sim T^2}$ for $\mathrm{T<T^*}$ because it creates a residual density of states for excitations at zero-energy. The curves in this figure are calculated from the theory of Hirschfeld and Goldenfeld~\cite{Hirschfeld1993} using the $\mathrm{T_c}$ values measured in Figure~\ref{tcplot}. For each $\mathrm{c}$ value and the ratio of $\mathrm{T_c}$ for sample $\mathrm{(X,Y)}$ to the $\mathrm{T_c}$ of pure YBCO, the crossover temperature $\mathrm{T^*}$ is calculated. The horizontal line at $\mathrm{26\ K}$ in the figure is the maximum temperature of the penetration depth experiment. Since $\mathrm{\lambda\sim T^2}$ for the $\mathrm{(0.36,0.16)}$ and $\mathrm{(0.26,0.26)}$ samples all the way up to the maximum temperature of the experiment, the largest possible scattering strength is $\mathrm{c\approx 0.55}$ in order that $\mathrm{T^*>26\ K}$. Since the Ca and Ce atoms substitute at Y sites that reside out of the $\mathrm{CuO_2}$ planes in YBCO, the scattering is weak (Born) scattering~\cite{Born2009}. Born scattering implies $\mathrm{c\gg 1}$. Hence, the observed $\mathrm{\lambda\sim T^2}$ for $\mathrm{(0.36,0.16)}$ and $\mathrm{(0.26,0.26)}$ cannot be explained by non-magnetic impurity scattering of D-wave counter-doped YBCO. Therefore, the penetration depth experiment in Figure~\ref{lambda} suggests an S-wave YBCO phase at high Ca and Ce counter-doping.}
\label{tstar}
\end{figure}


\section*{Methods}

\subsection*{Materials Synthesis}
The following chemicals were used for the synthesis of all the cuprate precursors: $\mathrm{BaCO_{3}}$ (99.8\% purity), $\mathrm{Y_{2}O_{3}}$ (99.99\%), $\mathrm{CuO}$ (99.9\%), $\mathrm{CeO_{2}}$ (99.9\%), and $\mathrm{CaCO_{3}}$ (99.8\%). The initial powders were mixed and homogenized in elemental ratios. Precursor powders were prepared by conventional multi-step solid-state reaction. The raw materials were mixed together and calcined in batches of several hundreds of grams. The calcination of YBCO-123 precursors were done in five steps at temperatures of 850 C, 870 C, 880 C, 890 C and 910 C with intermediate homogenization. Calcination of the counter-doped (Ca/Ce - YBCO) precursors had to be optimized to ensure correct phase formation and to avoid any liquid phase (Ba-Cu-O) loss. The final process was a function of the doping level. In all cases they include 2 low temperature calcinations in a powder form (850 C, 870 C), two calcinations in pre-pressed form (890 C, 910 C; 40g pellets, diameter 28 mm, uniaxial pressing) and several sinterings at 910 C (40g pellets, diameter 28 mm, uniaxial pressing + cold isostatic pressing). The progress of calcination was monitored by x-ray diffraction (XRD) to obtain maximal phase purity, as shown in Extended Table~\ref{phasecomp}. An x-ray Rietveld analysis was performed on \YXY\ with (X,Y) = (0.26, 0.13) and (0.32, 0.16) to ensure that the Ca and Ce atoms substituted onto the Y sites (see Extended Table~\ref{rietveld}).

For the (0.26, 0.26), (0.29, 0.29), and (0.32, 0.32) samples,
raw materials were mixed into a batch of 1 kg and milled using a vibratory disc mill. First, the powder was calcined at 850 °C and 870 °C, with a temperature ramp of 12h and a hold of 72h at ambient atmosphere. The powder was then uniaxially pressed into pellets (40g, 32mm diameter) with a force of 6 tons and subsequently pressed isostatically at 300 MPa. The pellets were sintered 4 times (once at 890 C and 3 times at 910 C with a ramp of 12h and hold of 48h for all sintering cycles). The pellets were crushed and milled between the sintering cycles to achieve high homogeneity, followed by the same pressing process.

BSE-SEM and EDS were performed on the (0.26, 0.26) sample.
The morphology was investigated using SEM with a FEG electron source (Tescan Lyra dual beam microscope, TESCAN Brno, s.r.o.). Elemental composition and mapping were performed using an EDS analyzer (X-MaxN) with a 20 $\mathrm{mm^2}$ SDD detector (Oxford Instruments, High Wycombe, UK) and AZtecEnergy software. Elemental maps were measured 5 times with an acquisition time of 100 sec. To conduct the measurements, the samples were placed on a carbon conductive tape and a BSE detector was used to obtain the SEM photography. The acquisition time for each photograph was set to 22 sec. SEM and SEM-EDS measurements were carried out using a 15 kV electron beam. These measurements were done on several locations with magnification from 5,000x to 30,000x, yielding the same data regardless of the location and magnification. 

Poly-crystalline cylindrical \YXY\ pellets with diameter 15 mm and height 6 mm for (X,Y) = (0.0, 0.0), (0.13, 0.0), (0.13, 0.13), (0.26, 0.13), (0.32, 0.16), (0.36, 0.16), (0.26, 0.26), (0.29, 0.29), and (0.32, 0.32) were synthesized.
Further annealing at low temperatures ($\mathrm{T<400\ C}$) was performed in the gases Air, Oxygen, Ozone, Argon, and Hydrogen in order to change the doping in the grains and grain boundaries. Over 60 different anneals were performed on over 100 samples during the course of this work. The length of the anneals varied from a few hours to several weeks. We found that every single sample we checked had a superconducting transition. This observation shows that superconductivity in Ca and Ce doped \YXY\ is remarkably robust. The room-temperature thermopower was measured for each pellet after the low-temperature anneals to determine the $\mathrm{CuO_2}$ plane hole doping level by applying the thermopower relation between thermopower and doping discovered by Tallon et~al.~\cite{Presland1991,Obertelli1992}. We found the maximum $\mathrm{T_c}$ occurs at the thermopower ($\mathrm{\approx +2\ \mu V/K}$) predicted by this relation.

\subsection*{Description of the Experiments}

The superconducting transition temperature, ${\mathrm T_c}$, and the temperature dependence of the penetration depth, $\mathrm{\lambda}$, were measured by the change in inductance of a coil placed on the sample at a fixed frequency of 500 kHz. For the $\mathrm{T_c}$ values in Figure~\ref{tcplot}, we made a 40-turn pancake coil with inner diameter of 5 mm and outer diameter of 8 mm. For $\mathrm{\lambda}$ shown in Figure~\ref{lambda}, we made a 100-turn pancake coil with inner diameter of 5 mm and outer diameter of 15 mm. The room-temperature inductance of the 40-turn and 100-turn coils at 100 kHz is $\mathrm{9.9\ \mu H}$ and $\mathrm{81.9\ \mu H}$, respectively.

Both coils are made from non-magnetic Phosphor-Bronze (CuSnP alloy) magnet wire with diameter 150 ${\mathrm{\mu m}}$. We used Phosphor-Bronze because it has a large resistivity ($\mathrm{\sim 8.9\ \mu\Omega}$-cm) that changes less than 0.2 \% from zero-temperature up to 26 K. Its skin depth at 500 kHz is $\mathrm{\sim 225\ \mu m}$.
The skin-depth is larger than the diameter of the wire. Hence, changes in the coil inductance due temperature dependent skin-depth changes can be neglected.

There is a redistribution of the current density in the cross-section of the wire due to the proximity effect arising from the diamagnetic screening currents in the superconducting sample. This effect pulls the current distribution in the coil wire towards the sample. We find that it increases the coil resistance by $\mathrm{\sim 80\ \%}$ between the normal state and the low temperature superconducting state leading to a $\mathrm{\sim 1-(1/1.8)=44\ \%}$ reduction in the cross-sectional area of the Phosphor-Bronze wire that carries current.
The important point is that, while this current distribution is different from the normal state current distribution, it changes the derivative of the coil self-inductance with penetration depth, $\mathrm{dL/d\lambda}$, negligibly over the temperature range of the $\mathrm{\lambda}$ experiment~\cite{Supplement}. Hence, it does not change the shape of the $L$ curves in Figure~\ref{lambda}.

A Cu block with resistive heating elements inside was mounted on the cold head using Silver paste as the thermal contact. A $\mathrm{Cernox^{TM}}$ semiconducting thermometer was mounted on the Cu block. The samples were pressed onto the Cu block using Apiezon-N grease.

The coil was kept in contact with the sample by the force from two small springs. The force from the springs was adjusted such that there was enough force to keep the coil on the sample while maintaining a small enough coil-to-sample force so that the thermal resistance between the coil and sample was large. This adjustment reduces uncertainties in the sample temperature arising from its contact with the coil that may be at a different temperature.
Estimates of the thermal diffusivity of the samples and the thermal resistance between the coil and sample can be found in the Supplement~\cite{Supplement}.

All our quoted $\mathrm{T_c}$ values are the onset $\mathrm{T_c}$ rather than the midpoint $\mathrm{T_c}$ because $\mathrm{T_c(onset)}$ was more consistent with the measured room-temperature thermopower. $\mathrm{T_c(midpoint)}$ was found to be sensitive to the nature of the low temperature annealing. Hence, the results in Figure~\ref{tcplot} are the maximum onset $\mathrm{T_c}$ for each (X,Y) = (Ca,Ce) doping.

In Figure~\ref{lambda}, the change in inductance, L, with temperature is plotted rather than the change in $\mathrm{\lambda}$ with temperature because the constant derivative, $\mathrm{dL/d\lambda}$, is known only approximately due to uncertainties in the coil-to-sample distance. Since the shape of the change in L is the same as the shape of the change in $\mathrm{\lambda}$, the gap symmetry question can be answered without knowing $\mathrm{dL/d\lambda}$. From microwave measurements of the change in the penetration depth~\cite{Hardy1993} of YBCO and the corresponding data in Figure~\ref{lambda}, $\mathrm{dL/d\lambda\approx 2.0\ nH/nm}$. A schematic of the experiment and the details of the method to extract the coil L from reflection coefficient measurements is in the Supplement~\cite{Supplement}.

The change in the penetration depth (measured as a change in inductance, L) was obtained for many samples. The L was measured as the sample was cooled down from 26 K to 4 K and then reheated back to 26 K. If the hysteresis in L was large, the run was rejected. The quoted L values average the cool down and warm up sweeps. Most of these samples had changes in inductance over the temperature range of $\mathrm{4-26\ K}$ that was a few times larger than the $\mathrm{\sim 20\ nH}$ change found for pure \YBCO\ (see Figure~\ref{lambda}). These samples were also rejected, despite having relatively sharp $\mathrm{T_c}$ transitions, because it was possible that a small fraction of grains inside the sample had $\mathrm{T_c < 26\ K}$, leading to an increase in inductance with temperature due to superconductor-normal phase transitions during the experiment.

The two additional issues that could affect the measurement are coil heating and the effect of the temperature dependence of the critical-current density of the Josephson junctions formed at the grain boundaries, $\mathrm{J_c(Josephson)}$. These two effects may lead to non-intrinsic temperature changes in the total inductance. All $\mathrm{\lambda}$ experiments were done at transmit power of $\mathrm{-40\ dBm=0.1\ \mu W}$. Approximately half of this power was absorbed and the remainder was reflected back to the Vector Network Analyzer (VNA). Additional experiments were run at $\mathrm{-50\ dBm=0.01\ \mu W}$ and $\mathrm{-30\ dBm=1\ \mu W}$. No change in the measured inductance was found between the $\mathrm{-40\ dBm}$ and $\mathrm{-50\ dBm}$ runs, indicating that the superconducting shielding currents in the sample were less than $\mathrm{J_c(Josephson)}$ and that there was negligible coil heating. The Supplement~\cite{Supplement} estimates possible errors.
The Supplement~\cite{Supplement} also describes the data acquisition with the VNA and our procedure for calculating the error bars for L in Figure~\ref{lambda}.

The Point-Contact-Andreev-Reflection (PCAR) data were obtained using a Cu tip that was attached to a rod driven by a micrometer. Measurements of the current-voltage (I-V) and differential conductance (dI/dV) characteristics were made using a conventional four-terminal probe arrangement with the conductance data obtained using a standard ac lock-in technique at a frequency of 10 kHz. The point contacts and samples were immersed in a liquid helium bath at 4.2 K. Further details of the measurement technique can be found in reference~\cite{Soulen1998}.

\section*{Acknowledgments}
We thank Thomas E. Sutto for suggesting Ce as a potential $+4$ oxidation state impurity atom that will reside at the Y site. This work was partially funded by the Office of Naval Research under Contract Number N00014-18-1-2679.

\section*{Competing Interests}
The authors declare they have no competing financial interests.

\section*{Author Contributions}
J.T.-K. and C.A.M. conceived the project and designed experiments for thermopower and transition temperature. 
J.T.-K performed the room-temperature thermopower measurements, devised and performed the penetration depth experiments, devised the
 coax-correction and current-distribution correction protocols described in the supplement, and did all the theory calculations.
Additional annealing at lower temperatures was done by J.T.-K. and C.A.M. They also wrote the paper.
T.H. synthesized the materials and contributed to the interpretation of the XRD, SEM, and EDX data.
M.L. performed the XRD, SEM, and EDX measurements, and did the materials characterization. M.L. also contributed to the materials characterization between sinters.
M.S.O. designed and performed the PCAR experiment.

\section*{Correspondence}
Correspondence and requests for materials should be addressed to J.T.-K. (email: jamil@caltech.edu).

\end{document}


\title{Supplementary Information: Potential Major Improvement in Superconductors for High-Field Magnets}

\author{Jamil Tahir-Kheli}\email{jamil@caltech.edu}
\affiliation{\Caltech}

\author{Tom\'a\v s Hl\'asek}
\affiliation{\CAN}
\affiliation{\Tomas}

\author{Michal Lojka}
\affiliation{\Tomas}
\affiliation{\CAN}

\author{Michael S. Osofsky}
\affiliation{\Towson}

\author{Carver A. Mead}\email{carver@caltech.edu}
\affiliation{\Caltech}

\maketitle

\section{Extracting the impedance from the twisted-coaxial cable experiment}

\FloatBarrier

The goal of the experiment is to obtain the impedance of the coil/sample as a function of temperature from the complex reflection coefficient measured at the Vector Network Analyzer (VNA) at the chosen fixed frequency. The imaginary part of the impedance is $\omega L$ where $\omega$ is the angular frequency and $L$ is the desired inductance. The angular frequency is $\omega=2\pi f$, where $f$ is the applied frequency. We used $f=500\ \mathrm{kHz}$. The change in $L$ with temperature is proportional to the change in the penetration depth.

If there was no coaxial cable between the VNA and the coil/sample, then 
a measured reflection coefficient of the coil/sample of $\Gamma$ translates into a coil/sample impedance, $Z$,

\begin{equation}
\label{seq1}
Z=Z_{0}\left(\frac{1+\Gamma}{1-\Gamma}\right),
\end{equation}

\noindent where $Z_0$ is the impedance of the VNA.
Therefore, a short ($Z=0$) has $\Gamma=-1$, an open ($Z=+\infty$) has $\Gamma=+1$, and $Z=Z_0$ leads to $\Gamma=0$.
In this paper, the HP8753D VNA we used has $Z_0 =50$ Ohms.

When the coil/sample is connected to the VNA through a coaxial cable, there is attenuation and a phase change to the reflection coefficient arising from its passage through the coax. The  measured reflection coefficient, $\Gamma_M$, must be corrected to obtain the desired coil/sample reflection, $\Gamma$. Once $\Gamma$ is calculated from $\Gamma_M$, then the coil/sample impedance is obtained using equation~\ref{seq1}.

If the experiment is performed at a fixed temperature, then the effect of the coax on the reflection coefficient could be removed from the measurement by using the standard Short, Open, Load (SOL) calibration method.

In an SOL calibration, three measurements are done. They are the reflection coefficients with the coax shorted at the position of the coil/sample, $\Gamma_S$, the coax left open, $\Gamma_O$, and with a known impedance load, $\Gamma_L$. When the coil/sample is connected to the end of the coax, the relation between measured reflection coefficient, $\Gamma_M$, and the coil/sample $\Gamma$ is

\begin{equation}
\label{seq2}
\Gamma_M = E_{11} + \frac{E_{12}\Gamma}{1-E_{22}\Gamma},
\end{equation}

\noindent where the coefficients $E_{11}$, $E_{12}$, and $E_{22}$ are

\begin{equation}
\label{seq3}
E_{11}=\Gamma_L,\ \ E_{12}=\frac{2(\Gamma_0-\Gamma_L)(\Gamma_L-\Gamma_S)}
                                {\Gamma_0-\Gamma_S},
\ \ E_{22}=\frac{\Gamma_O-2\Gamma_L+\Gamma_S}{\Gamma_O-\Gamma_S},
\end{equation}

\noindent leading to a simple expression for the coil/sample impedance as a function of $\Gamma_S$, $\Gamma_O$, and $\Gamma_L$

\begin{equation}
\label{seq4}
Z = Z_0 \left(\frac{\Gamma_L-\Gamma_O}{\Gamma_L-\Gamma_S}\right)
        \left(\frac{\Gamma_M-\Gamma_S}{\Gamma_M-\Gamma_O}\right).
\end{equation}

An SOL calibration for an experiment where the temperature of the sample varies has two diffculties. First, it requires a knowledge of the Load impedance at all temperatures. Second, a relay connected to the coax that toggles between Short, Open, Load, and the coil/sample is necessary. The relay must be placed as close to the coil/sample and Load impedance material as possible in order to reduce additional temperature-dependent reflections, phase shifts, and attenuation in any leads connecting the relay to the Short, Open, Load, and coil/sample circuits.

These two problems are eliminated by twisting two coaxes of equal length together and wiring up the experiment as shown in Supplementary Figure~\ref{coaxexp}. In this arrangement, the relay is outside the cryostat at the fixed lab temperature. It toggles between three configurations, Short, Open, and Load. Here, the Load is $\mathrm{50\ \Omega}$ at the lab temperature. The coil/sample is always connected to the ends of both coaxes inside the cryostat. By twisting the coaxes together, the temperature gradients on both coaxes are identical. Since the wavelength of the electrical waves in the coaxes is $\mathrm{\sim 400\ m}$ and each coax is $\mathrm{\sim 1\ m}$ in length, the coaxes can be modeled as transmission lines with an effective complex impedance, $Z_{coax}$, an effective length, $\mathrm{l_{coax}}$, and an effective complex attenuation, $A$. The complex attenuation includes the change in the amplitude of the wave traversing the coax and its phase shift over its length. $Z_{coax}$ and $A$ are unknown functions of the temperature of the coil/sample. They arise from the complicated thermal gradients over the lengths of the coaxes that is at lab temperature at one end and the coil temperature at the other end.

A single reflection coefficient measurement leads to a complex reflection, or two real numbers. The total number of unknowns at a given coil/sample temperature are two from the desired complex coil/sample impedance, $Z_S$, two from the unknown coax complex impedance, $Z_{coax}$, and two from the coax attenuation, $A$, or $2+2+2=6$.

By toggling through Short, Open, and Load measurements at the relay leads to three complex reflection coefficients, or 6 real numbers. Therefore, the effect of the unknown thermal gradients on the coax can be removed to obtain the coil/sample impedance, $Z_{S}$.

The expression for the measured reflection coefficient as a function of the uknown parameters, $Z_{S}$, $Z_{coax}$, and $A$, is obtained by solving for the transmitted and reflection waves at the relay and the junction with the coil/sample. There is a conservation of current condition and equality of voltage condition at each junction in the experiment. These equations can be solved for the reflection coefficient, $R$, as a function of the impedance, $Z$, at the end of the relay,

\begin{equation}
\label{reflect}
R = A\left[
           \frac
   {-Z_{coax}(Z+Z_{coax}) + (2Z_{S}-Z_{coax})(Z-Z_{coax})A}
   {(2Z_{S}+Z_{coax})(Z+Z_{coax}) + Z_{coax}(Z-Z_{coax})A}
     \right].
\end{equation}

\noindent Physically, $A$ is the round-trip attenuation (dimensionless) in a single coax.

Defining the dimensionless parameters,

\begin{equation}
\label{abpars}
\alpha = \frac{2Z_S}{Z_{coax}},\ \ \ \
\beta = \frac{Z_L-Z_{coax}}{Z_L+Z_{coax}},
\end{equation}

\noindent leads to,

\begin{equation}
\label{reflect1}
R = A\left[\frac{-1 + (\alpha-1)\beta A}{(\alpha+1) + \beta A}\right].
\end{equation}

The reflection coefficients for a Short ($Z=0$), Open ($Z=+\infty$), and Load ($Z=Z_L$) are,

\begin{equation}
\label{rs}
R_S = A\left[\frac{-1 - (\alpha-1) A}{(\alpha+1) -  A}\right],
\end{equation}

\begin{equation}
\label{ro}
R_O = A\left[\frac{-1 + (\alpha-1) A}{(\alpha+1) +  A}\right],
\end{equation}

\begin{equation}
\label{rl}
R_L = A\left[\frac{-1 + (\alpha-1)\beta A}{(\alpha+1) + \beta A}\right].
\end{equation}

These relations lead to an expression for $\alpha$,

\begin{equation}
\label{alpha1}
\alpha^2=(-)\left[\frac{2+R_O-R_S}{R_O+R_S}\right]
            \left[\frac{2+\frac{1}{R_O}-\frac{1}{R_S}}
                       {\frac{1}{R_O}+\frac{1}{R_S}}\right].
\end{equation}

\noindent Once $\alpha$ is known, $A$ satisfies,

\begin{equation}
\label{atten}
A=(-)(\alpha+1)\left[\frac{R_O+R_S}{2+R_O-R_S}\right].
\end{equation}

\noindent $\beta$ can then be solved by,

\begin{equation}
\label{beta1}
\beta=\left[(\alpha-1)A-R_L\right]^{-1}
      \left[1+R_L\left(\frac{\alpha+1}{A}\right)\right].
\end{equation}

\noindent Using equation~\ref{abpars} leads to $Z_S$.

\begin{figure}[tbp]
\centering \includegraphics[width=130mm]{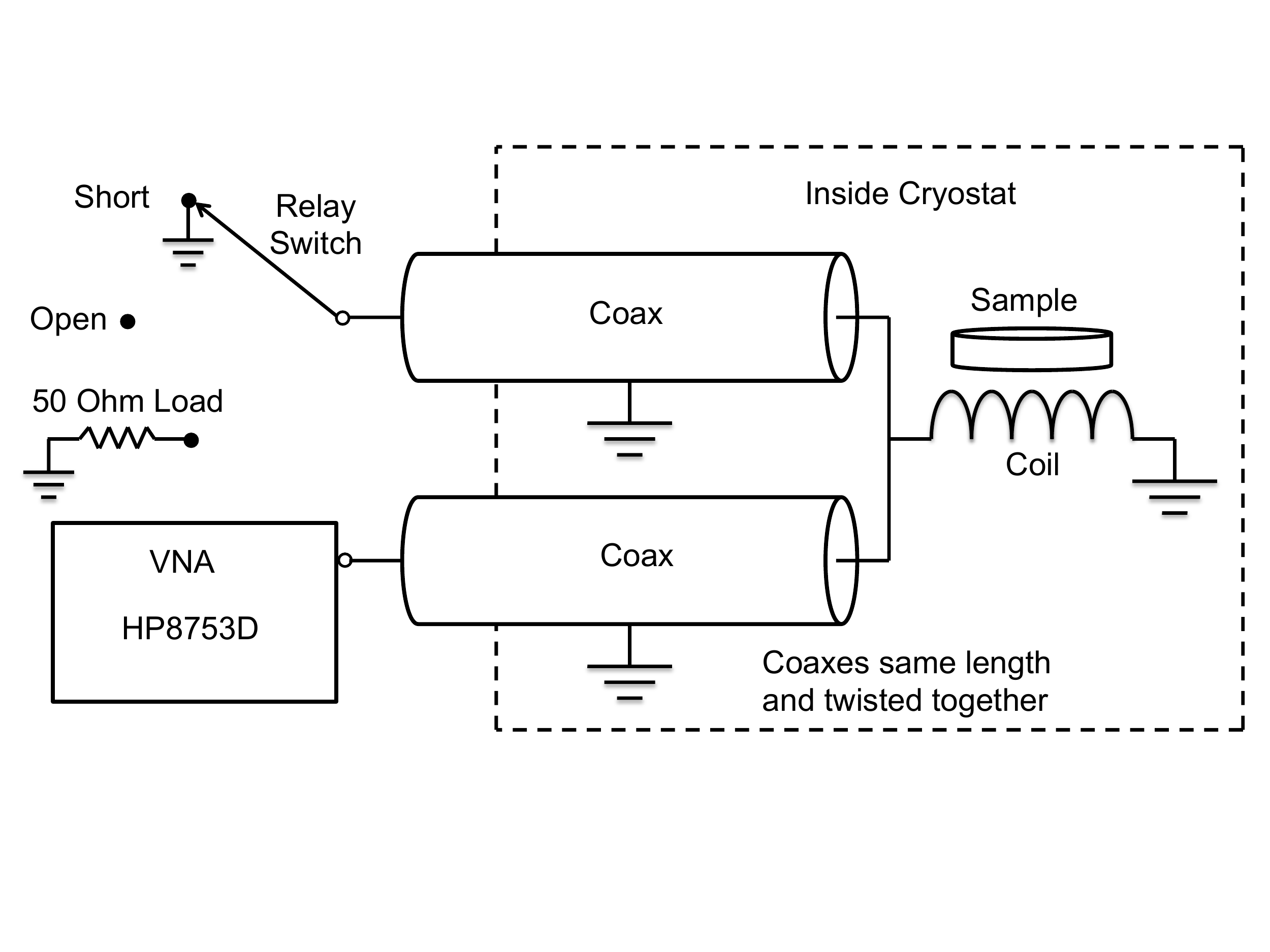}
\caption{
\textbf{Schematic Diagram of the $\mathbf{T_c}$ and penetration depth, $\mathbf{\lambda}$, experiments.}
The two coax cables are the same length and are twisted together so that the temperature gradients on the two coaxes are the same. A relay switches between Short, Open, and Load. The $\mathrm{50\ \Omega}$ load is the second port on the VNA.
}
\label{coaxexp}
\end{figure}

In the experiments, we cycle through Short, Open, and Load reflection coefficient measurements. Each measurement takes $\approx 6\ \mathrm{seconds}$ and $\approx 3\ \mathrm{seconds}$ of this time is spent acquiring the reflection coefficient. Since the Short, Open, and Load measurements occur at slightly different temperatures of the sample, linear interpolation is used to obtain the reflection coefficients for all three measurements at the same temperature.

For example, if a Short is measured at some temperature, then the Load measurement just prior to the Short and the Load measurement following the Open measurement right after the Short are used to obtain the Load reflection coefficient at the measured Short temperature.

\FloatBarrier

\section{Estimating the proximity and skin effects in the coil}

\begin{figure}[h]
\centering \includegraphics[width=100mm]{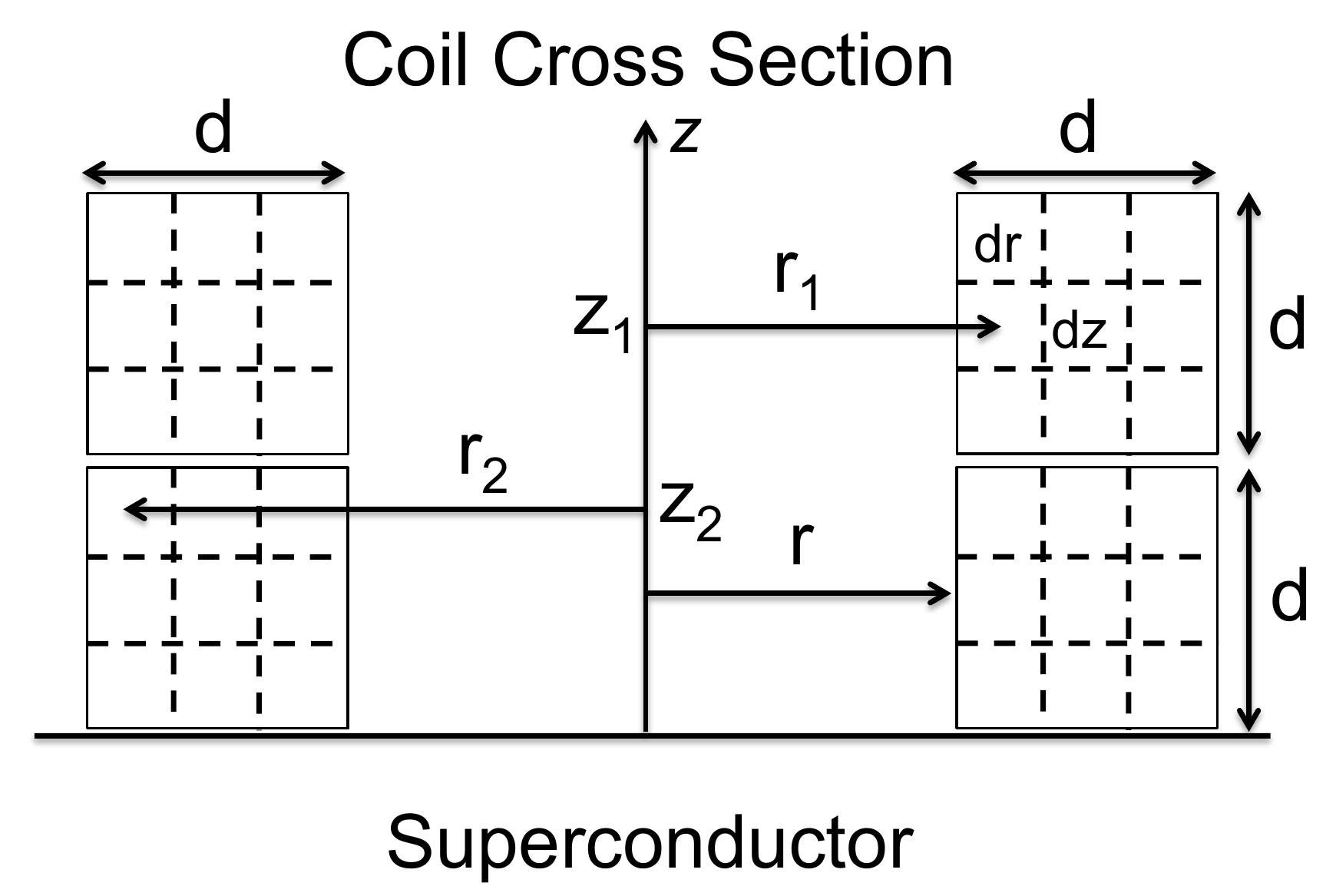}
\caption{
\textbf{Coil cross section on top of a superconducting surface.}
This schematic shows a case of two coil wire layers with a single coil turn in each layer. Each turn of the coil has radius $r$ and has rotational axis $z$. A $3\times 3$ array of discrete current loops is shown in each coil turn. They represent the discretization of the coil cross section for computing the current distribution inside each coil turn. The cross section of each discrete current loop is $dr\times dz$.}
\label{coilcross}
\end{figure}

Supplementary Figure~\ref{coilcross} shows the cross section of a coil with square cross section on the surface of a superconductor. In this Figure, the coil is formed from two coils with one on top of the other. An alternating current with angular frequency $\mathrm{\omega}$ is applied to the coil. There is a skin effect of each coil turn on itself that leads to the current distribution in the coil cross section residing predominantly on the coil surface.

There is a proximity effect on the current distribution inside a coil turn cross section arising from the parallel current in the neighboring coil turn. Since the current in the adjacent coil is in the same direction as the coil current, this proximity effect acts to push the current distributions in the two coil turns apart.

The superconducting screening current is in the opposite direction of the coil current direction. The proximity effect on the coil turns from this diamagnetic superconducting current acts to pull the coil current distribution closer to the superconducting surface. The competition between the skin effect and the two proximity effects above determines the current density distribution in the coils. Since the spatial distribution of the diamagnetic current varies with changes in the penetration depth, the current density distribution in the coils changes with changes in the penetration due to changes in temperature.

In our penetration depth experiment, we measure the change in the coil inductance, $L$, from 4 -- 26 K and assume that its change is proportional to the change in the penetration depth, $\mathrm{\lambda}$, with a proportionality constant that is temperature independent. This statement is equivalent to saying that $dL/d\lambda$ is constant during the experiment. Any change of the current distribution in the coil due to the proximity effect from the supercurrent will lead to a temperature dependent $dL/d\lambda$. The theoretical model in this section estimates $dL/d\lambda$ for our experiment and shows that changes in $dL/d\lambda$ are negligible. Hence, the shape of the $L$ curves shown in Figure~{4} is also the shape of the change in $\lambda$.

To determine the $\mathrm{T_c}$ values, shown in Figure~{2}, we measure the change in $L$ with $T$. What appears to be two phase transitions occurs for samples with broader $\mathrm{T_c}$, as seen in Supplementary Figure~\ref{tc26}. We show here that the two distinct slopes in $L$ versus $T$ are due to a large change of the coil current distribution towards the superconducting sample as the sample transitions from the normal state with a large metallic screening depth (on the order of millimeters) to a superconducting penetration depth (on the order of 100s of nanometers). Hence, our samples do not have two distinct $\mathrm{T_c}$ values.

Let $V_{\alpha}$ and $I_{\alpha}$ be the voltage and current in a current loop labeled by $\alpha$. Then

\begin{equation}
\label{veqn}
V_{\alpha} = R_{\alpha}I_{\alpha} + i\omega M_{\alpha,\beta}I_{\beta},
\end{equation}

\noindent where $R_{\alpha}$ is the resistance of current loop $\alpha$, $M_{\alpha,\beta}$ is the mutual inductance between current loops $\alpha$ and $\beta$, $I_{\beta}$ is the current of loop $\beta$, and there is an implicit sum over $\beta$ on the right-hand-side of the equation using the Einstein summation convention. The $i$ appearing before $\omega$ is the complex imaginary number $\sqrt{-1}$.

To explain our model, we use the coil shown in Supplementary Figure~\ref{coilcross} as a simple example that is easily generalized. First, we ignore
the superconductor in the coil arrangement of Figure~\ref{coilcross} by assuming the coil is far away from the superconductor. In this case, there are $9$ discrete voltages and currents for each coil turn for a total of $9+9=18$ unknown voltages and $18$ unknown current values. Hence, the total number of unknowns is $18+18=36$.

Since the voltage across each coil turn must be the same for each discrete current loop inside the coil turn, there are $9-1=8$ voltage equations for each coil turn for a total of $8+8=16$ voltage equations. They are

\begin{equation}
\label{vequal}
V_{1,t} = V_{2,t} = \ldots = V_{9,t}
\end{equation}

\noindent where $t$ is the coil turn number ($t=1,2$) and $V_{\alpha,t}$ is the voltage of the $\alpha$ current loop in coil turn $t$.

Since the total current in each coil turn must be equal, there are two additional current equations

\begin{equation}
\label{isum}
\sum_{\alpha}I_{\alpha,1} = \sum_{\alpha}I_{\alpha,2} = I
\end{equation}

\noindent where $I$ is the total coil current.

There are an additional $18$ equations connecting the voltages to the currents in the loops using equation~\ref{veqn} leading to a total of $18 + 16 + 2 = 36$ equations for $36$ unknowns. Therefore, given the coil current $I$, the voltage and current values for each loop can be obtained.

The remaining unknowns in equation~\ref{veqn} are the mutual inductances, $M_{\alpha,\beta}$ of the discrete current loops and their resistance. The mutual inductance, $M(r_1,r_2,z)$, between two coaxial current loops that are infinitely thin as a function of the radii of the two loops, $r_1$ and $r_2$, and their separation, $z$, is known to be the difference of two elliptic functions. For computational speed, we calculate $M(r_1,r_2,z)$ using the arithmetic-geometric mean method of King~\cite{King1921}.

We determine the mutual inductance between two distinct rectangular current loops by assuming their currents are concentrated at the center of each rectangle and then apply the above result for the mutual inductance between coaxial circles.

For the self-inductance, $M_{\alpha,\alpha}$, of rectangular current loop $\alpha$ with dimensions $dr$ and $dz$, such as the $r_1$ coil loop in Supplementary Figure~\ref{coilcross}, we use the value

\begin{equation}
\label{selfm}
M(r_1-\frac{1}{2}g_r,r_1+\frac{1}{2}g_r,g_z)
\end{equation}

\noindent where $g_r$ and $g_z$ are the geometric mean-distances of lines of length $\delta r$ and $\delta z$, respectively. The geometric mean-distance, $g(l)$, of a line of length $l$ is given by

\begin{equation}
\label{gmean}
g(l) = e^{-\frac{3}{2}}l\approx 0.223l.
\end{equation}

When the coil sits on top of the superconductor as shown in Supplementary Figure~\ref{coilcross}, equation~\ref{veqn} must include the effects of the superconducting screening current with a penetration depth of $\lambda$. We approximate the effect of the screening currents on the coil by choosing the boundary condition that the magnetic vector potential, $\mathbf{A}$, is zero at a depth of $\lambda$ below the superconductor surface, or $\mathbf{A}=0$.

\begin{figure}[tbp]
\centering \includegraphics[width=100mm]{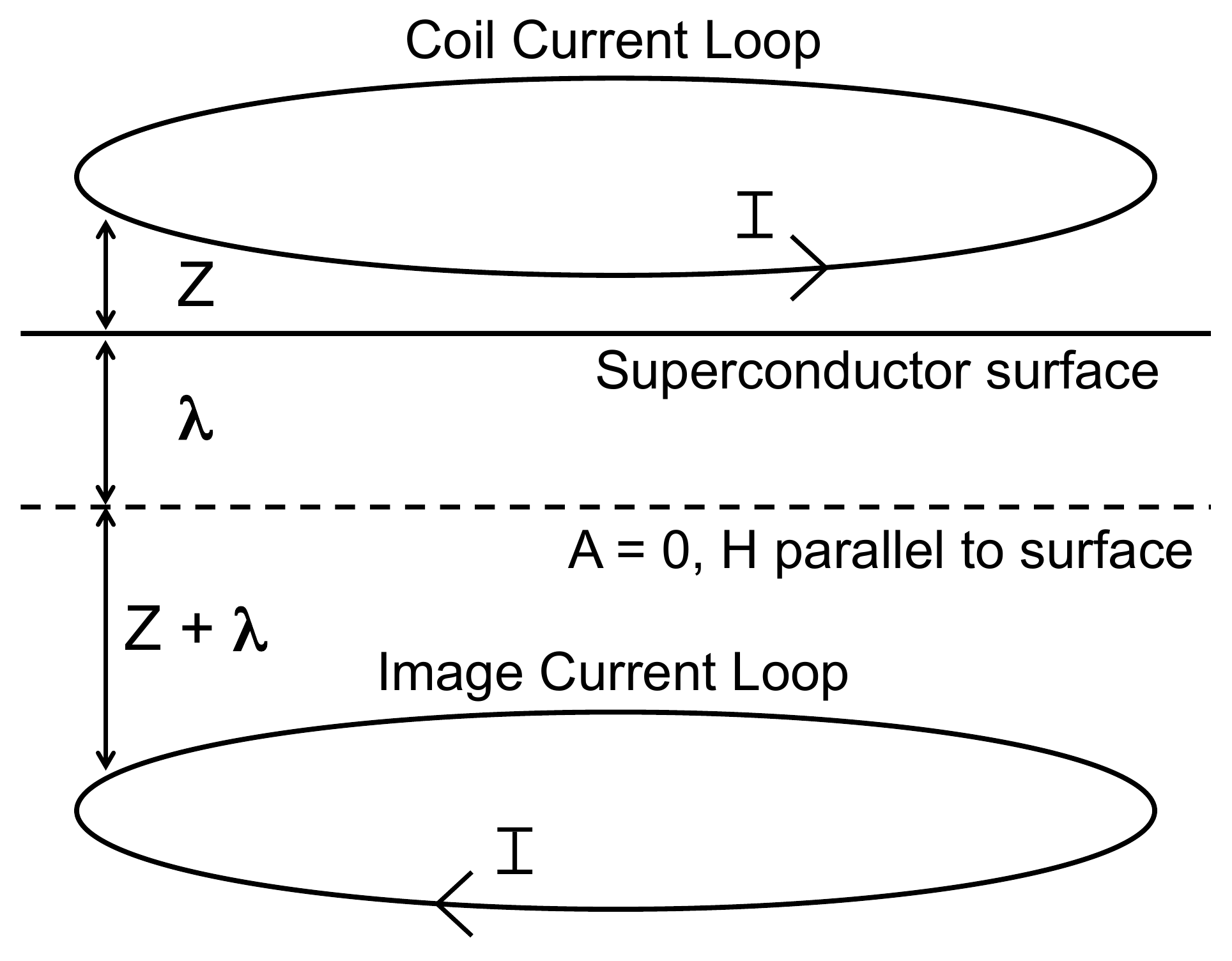}
\caption{
\textbf{A discrete current loop at height $Z$ above the superconducting surface and its image current at a depth $Z+\lambda$ below the superconducting surface.} When an equal and opposite current flows in the image current loop, the vector potential, $\mathbf{A}$, is zero at depth $\lambda$ below the surface. This condition, $\mathbf{A}=0$, at a depth $\lambda$ is the boundary condition for the diamagnetic screening current.
}
\label{imagecoil}
\end{figure}

For a single discrete current loop in the coil at a height $z$ above the superconductor surface, Supplementary Figure~\ref{imagecoil} shows an image coil at a depth of $z+\lambda$ below the superconductor surface. If the current in the image loop is in the opposite direction of the coil current loop, then $\mathbf{A}=0$ at a depth $\lambda$ below the surface. Hence, we approximate the diamagnetic response on the coil by image current loops carrying the current in the opposite direction as the corresponding current loop in the coil.
This image boundary condition, $\mathbf{A}=0$ at depth $\lambda$, is equivalent to the statement that there is no magnetic flux across depth $\lambda$.

The voltage equation~\ref{veqn} is modified to become

\begin{equation}
\label{veqns}
V_{\alpha} = R_{\alpha}I_{\alpha} +
i\omega\left[M_{\alpha,\beta}-M_{\alpha,\mathrm{Im}(\beta)}(\lambda)
\right]I_{\beta},
\end{equation}

\noindent where $M_{\alpha,\mathrm{Im}(\beta)}(\lambda)$ is the mutual inductance between the current loop $\alpha$ and the image current of current loop $\beta$ for penetration depth $\lambda$.

Solving equation~\ref{veqns} with the voltage relation equation~\ref{vequal} and the current sum equation~\ref{isum} includes the proximity effect of the supercurrent for each $\lambda$.

Ideally, we would like to model each coil turn with at least several hundred discrete coils in order to have a fine enough mesh to capture the current density changes in each coil turn. Since our small coil has $40$ turns and our large coil had $100$ turns, there are too many discrete currents loops to compute and the computation becomes intractable.

Hence, we simplified the model. We include $1,000$ and $2,000$ discrete current loops inside each coil turn in order to precisely model the current distribution in a coil turn, but we assume each coil turn has an effective number of turns, $n_{\mathrm{eff}}$, that approximately represents the $40$ and $100$ physical turns of our two coils.

An effective number of coil turns modifies equation~\ref{veqns} by multiplying the resistance terms by $n_{\mathrm{eff}}$ and the mutual inductance terms by $n_{\mathrm{eff}}^2$ to become

\begin{equation}
\label{veqnseff}
V_{\alpha} = n_{\mathrm{eff}}R_{\alpha}I_{\alpha} +
i\omega
n_{\mathrm{eff}}^2
\left[
M_{\alpha,\beta}-M_{\alpha,\mathrm{Im}(\beta)}(\lambda)
\right]I_{\beta},
\end{equation}

A smaller coil was used for the $T_c$ measurements and a larger coil was used for the penetration depth measurements. Here we model the smaller coil as 2 layers with one coil turn in each layer, as shown in Supplementary Figure~\ref{coilcross}. We discretize each coil turn with an $n_{r}\times n_{z}=10\times 100$ current loop grid where $n_r=10$ is the number of discrete loops in the radial direction and $n_z=100$ is the number of discrete loops in the z-axis direction (normal to the superconductor surface). The Figure shows a $3\times 3$ discrete loop configuration in each coil turn for simplicity. The radius of the coil, shown as $r$ in the Figure, is $3.75\ \mathrm{mm}$ and we choose the coil resistivity to be $\rho=0.8\rho_0$ where $\rho_0=8.59\ \mathrm{\mu\Omega}$-cm is the resistivity of phosphor-bronze at $T=4\ \mathrm{K}$.

The effective coil turns is $n_{\mathrm{eff}}=15$. Since there are two turns in two layers, the effective number of turns is $30$ compared to $40$ turns in the physical coil. The resistivity in this simplified model of the coil is chosen to be $20\%$ smaller than physical resistivity so that the calculated inductance and resistance, $L$ and $R$, respectively are close to the physically measured $L\approx 3.5\ \mathrm{\mu H}$ and $R\approx 4.5\ \mathrm{\Omega}$ at low-temperatures. With these parameters, the calculation gives $L\approx 16\ \mathrm{\mu H}$ and $R=3.3\ \mathrm{\Omega}$ at infinite penetration depth, or high-temperature. The measured inductance of the coil at $100\ \mathrm{kHz}$ at room-temperature is $9.9\ \mathrm{\mu H}$ and $R=3.3\ \mathrm{\Omega}$. Hence, our simplified computational model is a good representation of the physical coil.

\begin{figure}[tbp]
\centering \includegraphics[width=100mm]{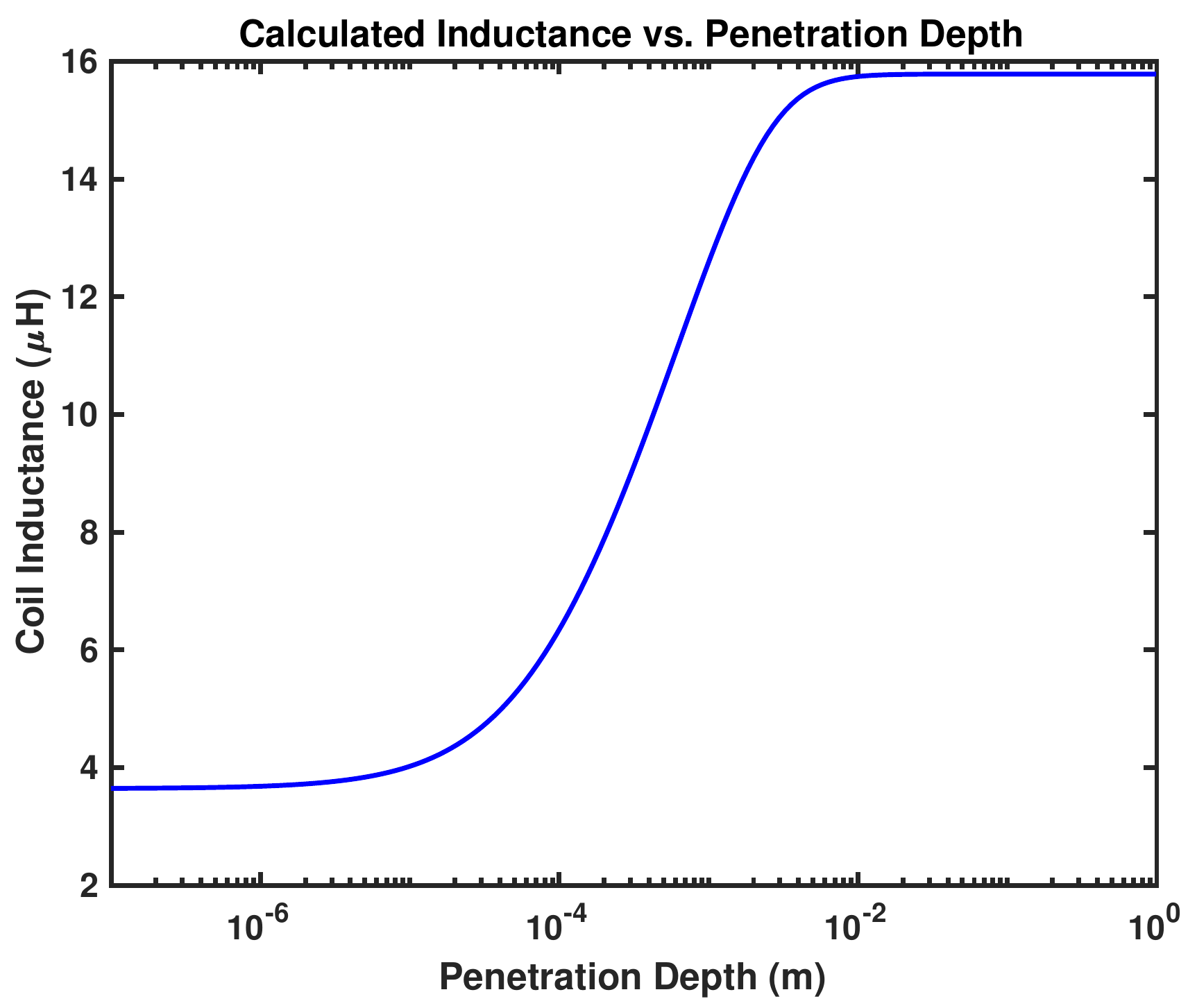}
\caption{
\textbf{Calculated coil inductance as a function of the superconducting penetration depth.}
The parameters used for this calculation are described in the text and are for the 40-turn coil.
}
\label{smallcoill}
\end{figure}

\begin{figure}[tbp]
\centering \includegraphics[width=100mm]{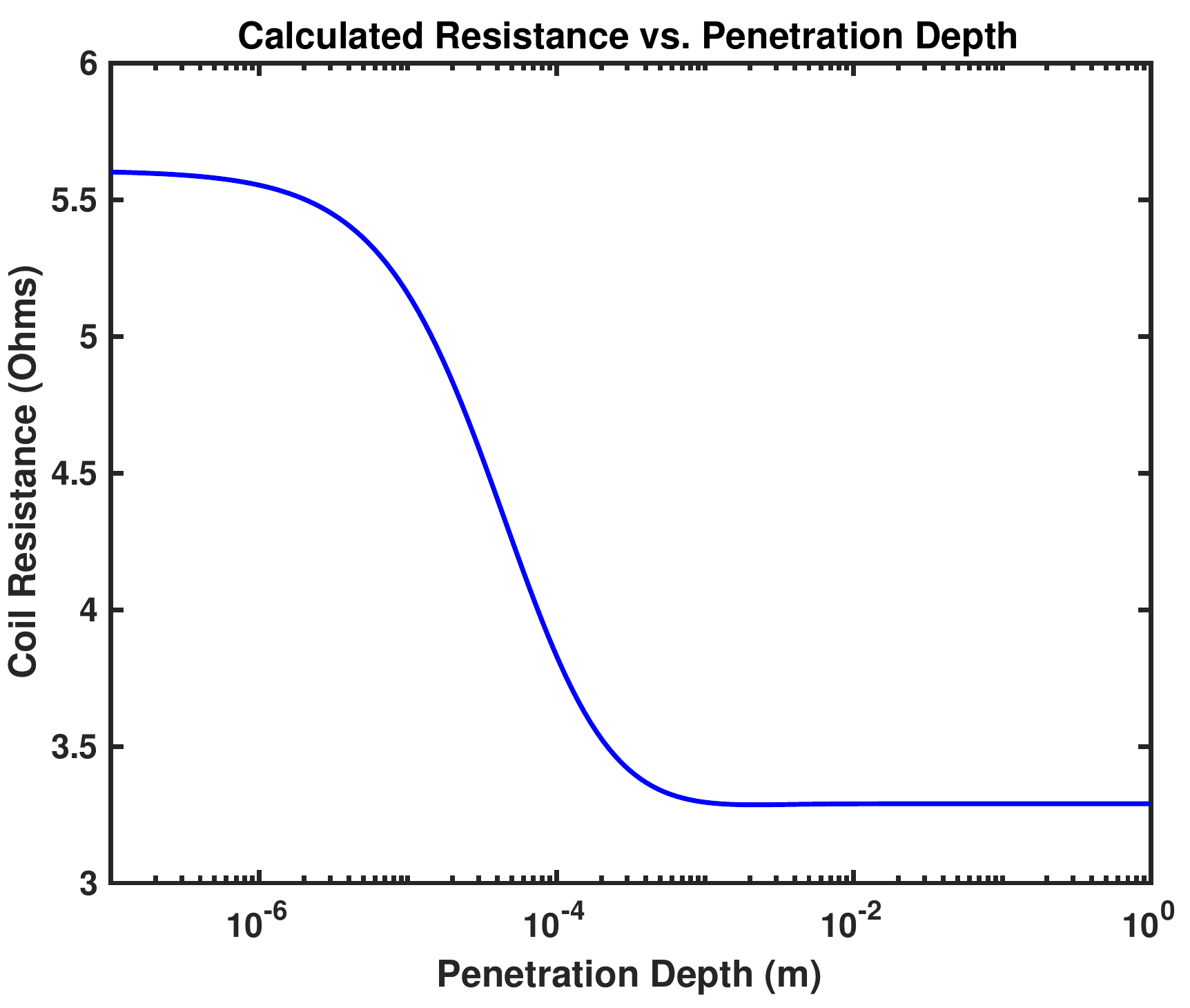}
\caption{
\textbf{Calculated coil resistance as a function of the superconducting penetration depth.}
The parameters used for this calculation are described in the text and are for the 40-turn coil. The increase in resistance at small penetration depth is due to the current density in the coil wires being pulled towards the supercurrent via the proximity effect.
}
\label{smallcoilr}
\end{figure}

Supplementary Figure~\ref{smallcoill} is the calculated $L$ as a function of the penetration depth and Supplementary Figure~\ref{smallcoilr} is the corresponding $R$. The $L$ plot is not surprising. There is a large screening effect due to the supercurrent as the penetration depth becomes small. The $R$ curve increases as the penetration depth decreases despite the resistivity of the coil remaining constant throughout the calculation. The $R$ increases because the proximity effect of the supercurrent is pulling the current distribution in the coil towards the superconductor surface. This concentration of current decreases the effective area of the current flow leading to an increase in resistance.

There are essentially two $R$ values. The low $R$ at infinite penetration depth (high-temperature) and the high $R$ value at small penetration depth (low-temperatures). Since the crossover of the penetration depth from infinity to microns occurs very close to $T_c$, there are ``two slopes" in the observed $L$ versus $T$ data, as seen in Supplementary Figure~\ref{tc26}. This calculation shows that the ``two $L$ slopes" does not occur from two distinct superconducting phases in our samples.

The 100-turn large coil has $L=81.9\ \mathrm{\mu H}$ and $R=20.2\ \mathrm{\Omega}$ at room-temperature at $100\ \mathrm{kHz}$. We model it by a single-layer coil with one turn. The radius $r=5\ \mathrm{mm}$, $n_r=10$, $n_z=200$, $n_{\mathrm{eff}}=100$, and $\rho=0.8\rho_0$.

\begin{figure}[tbp]
\centering \includegraphics[width=100mm]{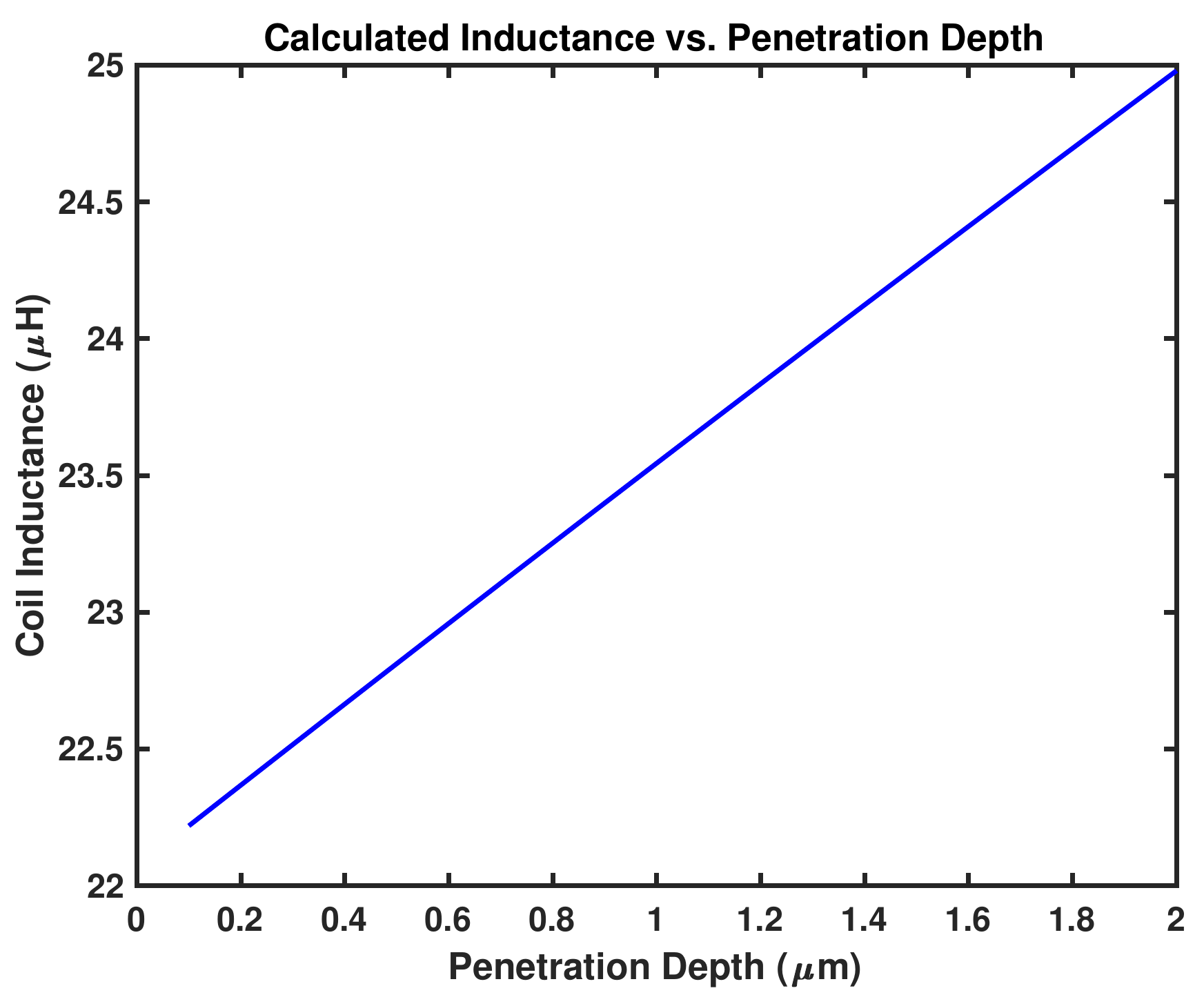}
\caption{
\textbf{Calculated change in inductance for the 100-turn coil as a function of the superconducting penetration depth between 100 nm and $\mathbf{2\ \mu m=2000\ nm}$.}
This calculation shows that the change in inductance is linear with penetration depth. Hence, $dL/d\lambda$ is constant over the range of the penetration depth experiment. Therefore, the shape of our measured change in $L$ curves in Figure 4 of the main text is identical to the shape of the change in penetration depth.
}
\label{largecoill}
\end{figure}

Supplementary Figure~\ref{largecoill} shows the change in $L$ versus $\lambda$ from $100\ \mathrm{nm}$ to $2\ \mathrm{microns}$. This computed $L$ contains the changes due to $\lambda$ and its proximity effect on the current distribution in the coil. The curve is linear. $dL/d\lambda$ is constant over the course of a penetration depth measurement. Hence, the measured shape of the $L$ curve is identical to the shape of the change in $\lambda$.

\FloatBarrier

\FloatBarrier

\section{Description of the $\mathrm{T_c}$ Experiment}

\subsection{$\mathrm{T_c}$ determination}

\FloatBarrier

The coil/sample inductance as a function of temperature for the two poly-crystalline \YXY\ samples with $\mathrm{(X,Y)=(0,0)\ and\ (0.26,0.26)}$ are shown in Supplementary Figure~\ref{tc00} and Supplementary Figure~\ref{tc26}. The $\mathrm{T_c}$ experiments were done with a $\sim10\ \mathrm{\mu H}$ coil. The dashed magenta line is a linear fit to the inductance just above the superconducting transition. The dashed red line is a linear fit to the inductance in the transition range of themperatures. The $\mathrm{T_c}$ used for Figure $1$ in the main text is the $\mathrm{T_c(onset)}$ defined by the intersection of the dashed magenta and the dashed red lines. As described in the main text, the $\mathrm{T_c(onset)}$ is used because it is stable to different low temperature sample anneals that lead to the same optimally doped room-temperature thermopower~\cite{Presland1991,Obertelli1992}.

\subsection{The two $\mathbf{T_c}$ slopes are due to the proximity effect rather than two superconducting phases}

There are two slopes in the superconducting transition plot in Supplementary Figure~\ref{tc26}. They appear for all of our samples when the $T_c$ transition width is at least several Kelvin. We believe they exist for all of our samples, but our warm up through the phase transition is too fast to resolve it for the samples with very sharp phase transtions.

The initial guess would be that there is two superconducting phases in this sample. This conclusion is incorrect. Supplementary Figure~\ref{smallcoilr} plots the calculated change in coil resistance as a function of the penetration depth. It shows that the proximity effect pulls the current density in the coil towards the sample, thereby increasing the coil resistance. Hence, the two slopes in the $L$ versus $T$ plots below $T_c$ are due to different higher and lower-temperature current density distributions in the coil rather than two distinct superconducting phases in the sample.

Typically, superconducting ac susceptibility experiments operate at much lower frequencies ($\mathrm{\sim 1-10\ kHz}$) compared to the frequency used here of $\mathrm{500\ kHz}$. The strength of the proximity effect is proportional to the applied frequency. In addition, the resistivity and diameter of the wire contribute. The frequency and choice of $150\ \mathrm{\mu m}$ phosphor-bronze wire for the coil material led to an observable current redistribution in Supplementary Figure~\ref{tc26}.

\begin{figure}[tbp]
\centering \includegraphics[width=100mm]{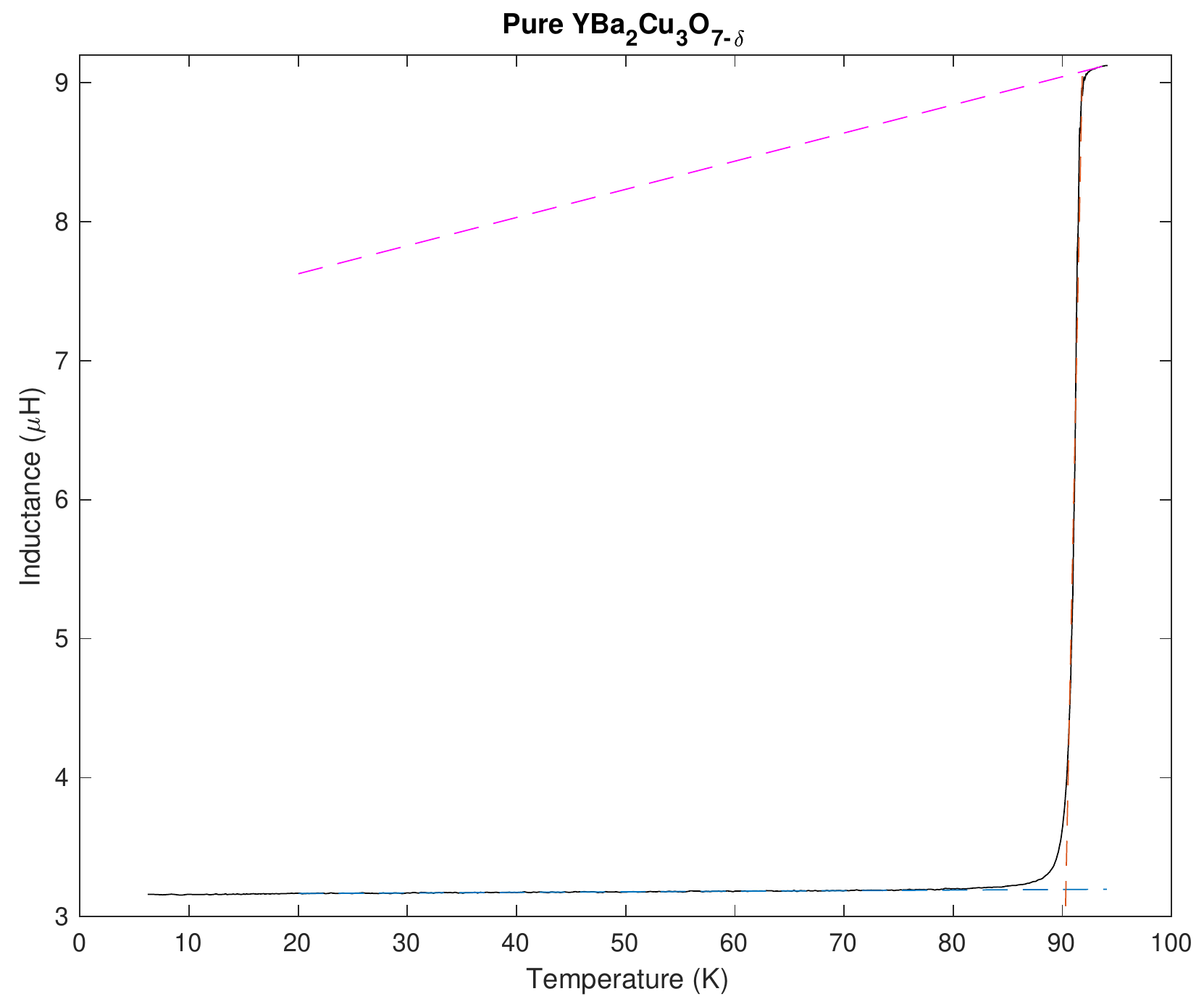}
\caption{
\textbf{$\mathbf{T_c}$ determination of pure
$\mathbf{YBa_{2}Cu_{3}O_{7-\delta}}$ from the change in inductance with temperature.}
The $\mathrm{T_c}$ is the onset $\mathrm{T_c}$ defined as the intersection of the red and magenta dash lines. Here, $\mathrm{T_c=91.8\ K}$.
}
\label{tc00}
\end{figure}

\begin{figure}[tbp]
\centering \includegraphics[width=100mm]{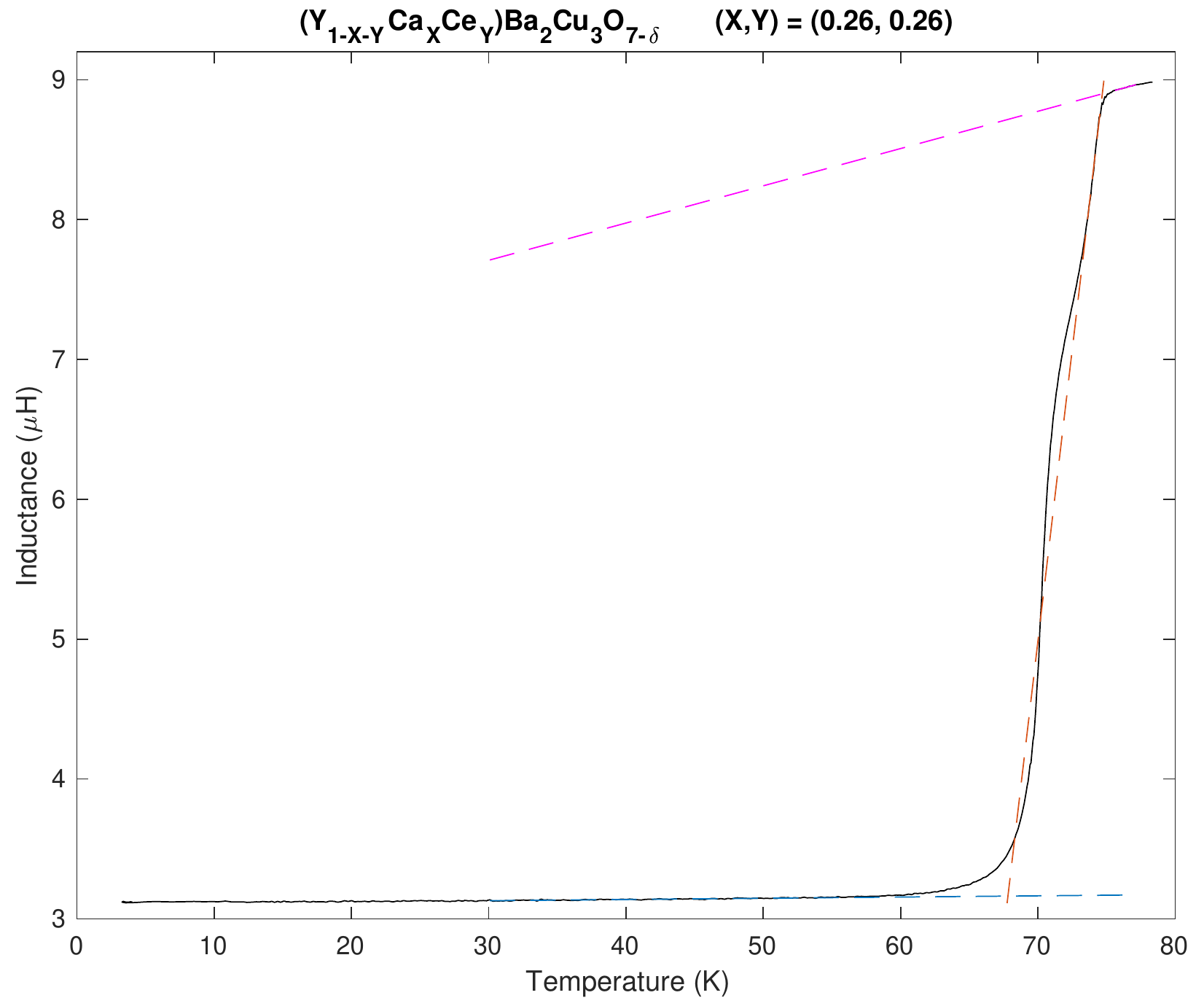}
\caption{
\textbf{$\mathbf{T_c}$ determination of
$\mathbf{(Y_{1-x-y}Ca_{x}Ce_{y})Ba_{2}Cu_{3}O_{7-\delta}}$
for (X,Y) = (0.26,0.26) from the change in inductance with temperature.}
The $\mathrm{T_c}$ is the onset $\mathrm{T_c}$ defined as the intersection of the red and magenta dash lines. Here, $\mathrm{T_c=74.7\ K}$. The change in slope during the phase transition (black line) is due to the onset of the proximity effect of the supercurrent in the sample pulling the coil current density towards the sample. It is not a second superconducting phase. The strength of the proximity is proportional to the applied frequency. At our frequency of $\mathrm{500\ kHz}$, the proximity effect leads to a noticable change in inductance as the coil current redistributes from the normal state distribution to the superconducting distribution.
}
\label{tc26}
\end{figure}

\FloatBarrier

\section{Calculating the $\mathrm{T_c}$ from Non-Magnetic Pair-Breaking Theory}

\FloatBarrier

\begin{figure}[tbp]
\centering \includegraphics[width=120mm]{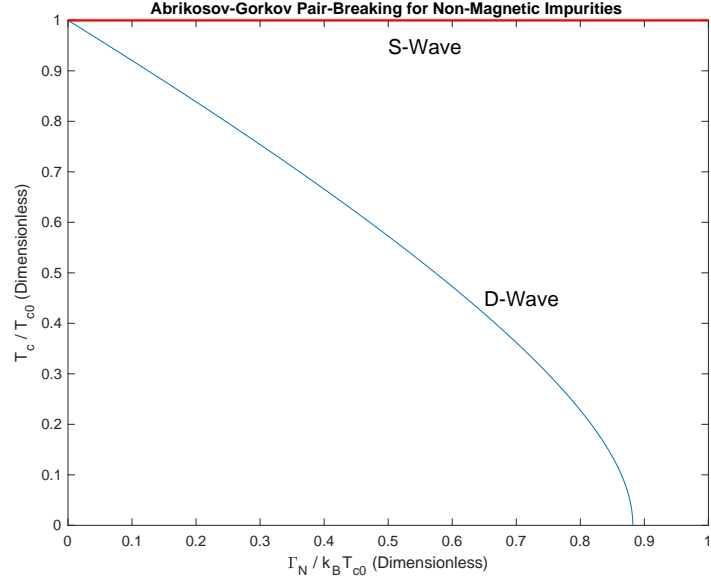}
\caption{
\textbf{The Abrikosov-Gorkov theory for the change in D-wave and S-wave superconductors as a function of the pair-breaking strength of non-magnetic impurities.}
 The x-axis and y-axis are normalized to $k_{B}T_{c0}$ and $T_{c0}$, respectively, where $k_B$ is Boltzmann's constant and $T_{c0}$ is the superconducting transtion temperature with no pair-breaking. The blue curve is the change in $T_c$ for a D-wave superconductor and the red curve is the $T_c$ change for an S-wave superconductor.
}
\label{AG1}
\end{figure}

Supplementary Figure~\ref{AG1} shows the drop in $\mathrm{T_c}$ as a function of the pair-breaking strength, $\mathrm{\Gamma_N}$. $\mathrm{T_{c0}}$ is the transition temperature of a pure undoped sample and $\mathrm{k_B}$ is Boltzmann's constant. This Figure was originally calculated by Abrikosov and Gorkov~\cite{Abrikosov1961} for the $\mathrm{T_c}$ change of an S-wave superconductor as a function of magnetic pair-breaking impurities. The Figure is also correct for D-wave gap symmetry pair-breaking by non-magnetic impurities~\cite{Hirschfeld1993}. Non-magnetic impurities have a very weak effect on the $T_c$ on an S-wave gap (known as Anderson's Theorem~\cite{Anderson1959}).

Since Ca and Ce atoms are non-magnetic in \YBCO, this curve predicts the drop in $\mathrm{T_c}$ to be expected in our samples if the superconducting gap remained D-wave throughout the full Ca/Ce counter-doping range from $(0,0)$ to $(0.32,0.32)$. The red and green points in Figure 2 of the main paper are obtained from this Abrikosov-Gorkov pair-breaking curve as described below.

For the red points, called ``simple pair-breaking theory" in this Figure, the $\mathrm{T_c}$ values for pure \YBCO\ and \YXY\ with $\mathrm{(X,Y)=(0.13,0.0)}$ are obtained experimentally. The pure \YBCO\ $\mathrm{T_c}$ is set equal to $\mathrm{T_{c0}}$ in Supplementary Figure~\ref{AG1} and the dimensionless ratio, $T_{c}(0.13,0.0)/T_{c0}\approx 0.92$ is calculated. The corresponding dimensionless pair-breaking value on the x-axis of Supplementary Figure~\ref{AG1} for a value of $0.92$ on the y-axis is $\approx 0.1$. In this simple pair-breaking model, we assume Ca and Ce impurities lead to exactly the same pair-breaking. Hence, we conclude that $13\%$ non-magnetic impurities leads to $\approx 0.1$ pair-breaking strength in dimensionless units.

For $(0.36,0.16)$ counter-doping, the ratio of the number of non-magnetic impurities ($0.36+0.16=0.52$) to $0.13$ non-magnetic impurites is $4$. Hence, the pair-breaking for $(0.36,0.16)$ is $4\times 0.1\approx 0.4$. The y-axis value on the Abrikosov-Gorkov pair-breaking curve that corresponds to an x-axis value of $\approx 0.4$ is $\approx 0.67$ leading to $T_c\approx 63\ \mathrm{K}$ for a $(0.36,0.16)$ sample, as shown on Figure 2. The complete red plot in Figure 2 can be obtained in a similar manner.

For the green plot in Figure 2, called ``dipole pair-breaking theory", we assume a Ca atom will be close to a Ce atom since Ca ($+2$ oxidation state) and Ce ($+4$) have $+1$ and $-1$ charges relative to Y ($+3$). Thus they form a ``dipole" pair-breaker. Hence, the $T_c$ of the $(0.13,0.13)$ sample leads to a pair-breaking magnitude ($\approx 0.18$) using the method described above. This value is not the same as twice the pair-breaking magnitude for $(0.13,0.0)$ obtained above ($\approx 0.2$). Using the pair-breaking value for ``singles" and ``dipoles", the pair-breaking value for $(0.36,0.16)$ can be broken into an $(0.16,0.16)$ dipole pair-breaker and an $0.16$ single pair-breaker leading to a pair-breaking strength of $\approx 0.35$ and a predicted $T_c\approx 65\ \mathrm{K}$. By this reasoning, the complete green plot in Figure 2 is obtained.

For both models of pair-breaking, the experimental data is not compatible with a D-wave gap symmetry for all of the samples. The observed saturation of $T_c$ at higher Ca and Ce counter-doping indicates that these samples are S-wave in character.

\FloatBarrier

\section{Resistivity and Thermal Expansion of the Coil}

\FloatBarrier

\subsection{Resistivity changes in the coil}

Since the phosphor-bronze coil is not at a fixed temperature over the course of the penetration depth measurement, the size of the inductance changes due to changes in the coil resistivity must be estimated. We present two arguments indicating that changes in the coil resistivity lead to a negligible change in the measured coil inductance.

First, low temperature experiments measuring the resistivity of phosphor-bronze find it to be constant within error bars over the 4--26 K range of the penetration depth experiment~\cite{Clark1970,Tuttle2010}. Second, since the $T$ dependent contribution to the resistivity of phosphor-bronze is due to phonons, large changes in its resistivity are not possbile at low temperatures (approximately $\mathrm{< 20\ K}$). If there is a contribution to the coil inductance arising from resistivity changes in the coil, it would appear above 20 K, and lead to an increase in the linear $T$ slope of $L$ for the D-wave samples, (X,Y) = $(0.0, 0.0)$ and $(0.13, 0.0)$, in Figure 4 of the main text. Since there is no evidence for a change in the linear $T$ slope of $L$ for these two D-wave samples, we conclude that resistivity changes in the phosphor-bronze coil are not affecting the results in Figure 4.

\subsection{Thermal expansion of the coil}

The mean thermal expansion, $\alpha(T)$, of phosphor-bronze at low temperatures is obtained from Table $\mathrm{21.7}$ in chapter $21$ page $20$ of reference~\cite{NIST-Cu1992}. We use the data in the left table because $\alpha(T)$ is available for $T=0\ \mathrm{K}$ and $T=30\ \mathrm{K}$. $\alpha(T)$ is defined as

\begin{equation}
\label{expansion}
\alpha(T) = \left[\frac{L(293\ K) - L(T)}{L(293\ K)}\right]
            \left[\frac{1}{293\ K - T}\right],
\end{equation}

\noindent where $L(T)$ is the length at temperature $T$ in Kelvin. Reference~\cite{NIST-Cu1992} gives
$\alpha(0\ \mathrm{K})=11.26\times 10^{-6}\ \mathrm{K}^{-1}$ and
$\alpha(30\ \mathrm{K})= 12.51\times 10^{-6}\ \mathrm{K}^{-1}$.
$L(293\ \mathrm{K})=150\ \mathrm{\mu m}$, the diameter of the phosphor-bronze wire, leading to

\begin{equation}
\label{size}
L(0\ \mathrm{K}) = 150\ \mathrm{\mu m} - 494.9\ \mathrm{nm},\ \ \ 
L(30\ \mathrm{K}) = 150\ \mathrm{\mu m} - 493.5\ \mathrm{nm}.
\end{equation}

Thus the change in coil-to-sample distance due to thermal expansion over the 4--26 K temperature range of the penetration depth experiment is less than 1.4 nm. Since $dL/d\lambda\sim 2\ \mathrm{nH}/\mathrm{nm}$, a change in inductance of
$<2.8\ \mathrm{nH}$ occurs. This value is approximately an order of magnitude smaller than the observed changes in inductance in Figure 4. Hence, the effect of thermal expansion of the phosphor-bronze wire on the coil-to-sample distance may be neglected.

A 1.4 nm change in 150 $\mu\mathrm{m}$ is a relative change of $\approx 10^{-5}$. A $10^{-5}$ contraction in the overall size of the pancake coil leads to a change in inductance $<10^{-5}L_0$ where $L_0\approx 20\ \mathrm{nH}$ is the magnitude of the coil inductance during the penetration depth experiment. Therefore, the change in $L$ due to temperature changes in the diamater of the coil is $< 0.2\ \mathrm{nH}$ and may be also be neglected.

\FloatBarrier

\section{Miscellaneous sources of error}

For the penetration depth experiments shown in Figure 4 of the main paper, the inductance of the coil at room-temperature is $\approx 82\ \mathrm{\mu H}$. Its magnitude at $4\ \mathrm{K}$ is $\approx 20\ \mathrm{\mu H}$. The magnitude of the inductance change in Figure 4 is a few 10s of nanoHenrys from $\mathrm{4-26\ K}$. Changes on the order of $\sim\mathrm{1\ nH}$ must be resolved in order to distinguish between a temperature dependence of $T$ or $T^2$. The relative change in inductance that needs to be resolved is $\sim\mathrm{5\times 10^{-5}}$. Therefore, all potential sources of error must be estimated.

\subsection{Errors due to oscillator drift}

The oscillator drift is a relative frequency error of $3\times 10^{-9}$ per day. The experimental data shown in Figure 4 is acquired in $\approx 1.5\ \mathrm{hours}$ for each sample leading to a relative frequency error of $1.85\times 10^{-10}$ during the run. The relative error in the coil inductance, $L$, is of the same order. The magnitude of the observed inductance is $L\sim 20\ \mathrm{\mu H}$. Hence, the error due to oscillator drift is $\sim 3.75\times 10^{-15}\ \mathrm{H}$. Since the observed change in $L$ during the experiment is $\sim 10\ \mathrm{nH}$, the oscillator drift error is negligible.

\subsection{Errors in the load impedance reference}

Since the conclusion regarding the gap symmetry from the penetration depth measurement is based on the shape of the curves in Figure 4 rather than their absolute magnitude, an absolute error in the load impedance is unimportant. What matters is the drift of the load impedance over the course of the experiment. We use the second port on our VNA as a stable load impedance reference. Its impedance is stable to better than $10^{-5}$ for a change in the lab temperature of $\pm 1$\ Kelvin. During the duration of the experiment, the lab temperature does not change by more than a few Kelvin. In this worst case scenario, a change in lab temperature of 5 Kelvin leads to a relative change in the load impedance of $5\times 10^{-5}$ and a change in inductance, $L$, of the same order. For $L\approx 20\ \mathrm{\mu H}$, the change in $L$ is $\sim 1\ \mathrm{nH}$ and may be neglected.

\subsection{Errors due to coil heating}

Heating of the coil occurs while current flows through it during the reflection coefficient measurements that determine the coil/sample impedance. This heating increases the resistivity of the phosphor-bronze wire comprising the coil leading to a redistribution of the current density in the coil. If the heating affect is sufficiently large, it will introduce an error in the extracted coil inductance. This heating effect is largest at the lowest temperatures because the specific heat of the coil wire decreases with temperature.

In order to eliminate coil heating errors, the experiments are performed at a transmit power of $-$40 dBm, or 0.1 $\mathrm{\mu W}$. Approximately 1/2 of this power is absorbed during the experiment with most of it being dissipated in the coil. The remaining power is reflected back to the VNA. If there was a change in inductance arising from coil heating, then the same experiment performed at $-$50 dBm, or 0.01 $\mathrm{\mu W=10\ nW}$, would lead to different inductance values. We have repeated several runs at $-$50 dBm to check for coil heating. The shape of the inductance curve versus $T$ and its magnitude are almost identical. Supplementary Figure~\ref{2626run} and Supplementary Figure~\ref{2626-50dBm-run} show the raw data for the inductance change with $T$ of the $(0.26,0.26)$ sample at power levels of $-40\ \mathrm{dBm}$ and $-50\ \mathrm{dBm}$. Supplementary Figure~\ref{2626-compare} is a comparison of the inductance changes of the two runs. We conclude coil heating is not a significant source of error in our measurements.

\subsection{Errors due to Josephson effects in grain boundaries}

The magnetic field from the current in the coil induces a superconducting screening current in the sample. The measured inductance of the coil is a weighted average of the London penetration depth, $\lambda$, of the single-crystal grains in the sample and the Josephson junction penetration depth at the grain boundaries, $\lambda_{\mathrm{J}}$. This weighted average, the effective penetration depth, $\lambda_{\mathrm{eff}}$ is what is measured by the coil inductance. If $\lambda_{\mathrm{J}}$ is $T$ independent, then the $T$ dpendence of $\lambda_{\mathrm{eff}}$ is the same as the desired $\lambda$.

To investigate the $T$ dependence of $\lambda_{\mathrm{J}}$, we compare two runs of a $(0.26,0.26)$ sample at $-40\ \mathrm{dBm}$ and $-50\ \mathrm{dBm}$ in Supplementary Figure~\ref{2626-compare}.
The current density in the sample is $\sqrt{10}=3.16$ times smaller for the $-50\ \mathrm{dBm}$ run compared to the $-40\ \mathrm{dBm}$ run. The $\approx1\ \mathrm{nH}$ difference between the higher and lower power levels is tiny on the scale of the changes in inductance of the experiment. Also, it does not change the $T^2$ shape of the curve. Hence, changes in inductance arising from any $T$ dependence of $\lambda_{\mathrm{J}}$ is small and does not alter the conclusions of this paper.

\subsection{Errors due to thermal contact between the coil and sample}

The sample is placed on a Cu block with Apeizon-N grease for good thermal contact. The Cu block is thermally connected to the coldest part (stage 2) of the cold head. The thermometer is thermally mounted on the Cu block. The two ends of the coil are soldered to the twisted coaxial cables that are thermally clamped to the $\sim\mathrm{30\ K}$ first stage of the cold head and then twisted around the cold head cylinder using Apiezon-N grease two-thirds of the way towards the cold head at stage 2 where the Cu block is located. Hence, the coil and the sample will not be at the same temperature.

If there is a large thermal contact (or low thermal resistance) between the coil and the sample, then the sample temperature may be different from the reading of the thermometer and the measured inductance will not correspond to the temperature reading of the thermometer. What is desired is a large thermal resistance between the coil and sample while keeping the coil on the sample.

The coil is held on the sample by the force of two springs. Since thermal resistance is proportional to the magnitude of the force pressing the coil to the sample, the coil is pressed onto the sample using the minimum possible force possible. We tightened the coil onto the sample with a large force (maximum spring compression) and then did experiments with reduced force until the experiments converged to a consistent result.

The strongest evidence that the thermal contact between the coil and sample is small is the observed linear $T$ penetration depth for the known D-wave samples, $(\mathrm{Ca},\mathrm{Ce})=(0.0,0.0)$ and $(0.13,0.0)$, as shown in Supplementary Figures~\ref{0000run} and ~\ref{1300run}. However, hysteresis is seen in the cool down versus warm up sweeps (blue versus red data) for the $\mathrm{(0.13,0.0)}$ and $\mathrm{(0.26,0.26)}$ samples as shown in Supplementary Figures~\ref{1300run} and \ref{2626run}, likely due to this effect. By averaging the cool down and warm up data, we have eliminated this effect to first order.

\subsection{Errors due to a temperature gradient on the sample}

The thermometer and the bottom of the sample are thermally connected to the Cu block on the cold head. The sample is a cylinder with a diameter of $15\ \mathrm{mm}$ and height of $6\ \mathrm{mm}$. The coil is placed on the top of the sample. If there is a temperature difference between the bottom and top of the sample, then the measured inductance will not be at the temperature indicated by the thermometer.

Since we slowly cool and then warm up the sample between 4--26 K while acquiring data, the rate for the sample to thermally equilibrate must be faster than the rate of the temperature sweep in order to eliminate thermal gradient errors. Hence, the thermal diffusivity of the samples must be estimated.

The thermal diffusivity is the ratio of the thermal conductivity (units of W/m-K) and the specific heat per volume (units of $\mathrm{J/m^3 K}$). It is given by $(1/3)v^2 \tau=(1/3)v\Lambda$, where $v$ is the velocity of the phonons, $\tau$ is the scattering time, and $\Lambda=v\tau$ is the scattering length.

At low-temperatures, the phonon velocity is due to long-wavelength acoustic modes. Hence, the mass difference of the Ca and Ce atoms compared to the Y atoms will not lead to any substantial change in acoustic phonon velocity. We may estimate the velocity to be on the order of $v\sim 10^{3}\ \mathrm{m/s}$.

The scattering length at low-temperatures is large for single crystals. For our poly-crystalline samples, the scattering length becomes the size of the single-grains, or $\Lambda\sim\ 10\ \mathrm{\mu m}$. Thus the thermal diffusivity is $\approx 0.33\times 10^{-2}\mathrm{m^2/s}$. The time to thermally equilibrate over a $6\ \mathrm{mm}$ sample height is $\approx\ 10^{-2}\ \mathrm{s}$. The temperature sweep rate of the penetration depth experiment is $\approx 2\ \mathrm{min/K}$. Therefore, the sample temperature is almost uniform during the experiment and any errors due to thermal gradients on the sample are negligible.

\subsection{Errors due to vortices potentially entering and leaving the sample}

If the applied magnetic field on the sample is larger than H$_{c1}$, then vortices can enter the sample and affect the inductance. H$_{c1}$ is $10-20\ \mathrm{mT}$ for YBCO.~\cite{Larbelestier2001} The applied magnetic field at the surface of the superconductor is $\mathrm{H}\sim 0.1\ \mathrm{A/m}$ leading to
$\mu_0\mathrm{H}\sim 10^{-4}\ \mathrm{mT}<<\mathrm{H}_{c1}$. Therefore, the sample remains in the Meissner phase. Vortices do not enter the sample.

\FloatBarrier

\section{Raw data plots of the penetration depth experiments}

\FloatBarrier

\begin{figure}[tbp]
\centering \includegraphics[width=100mm]{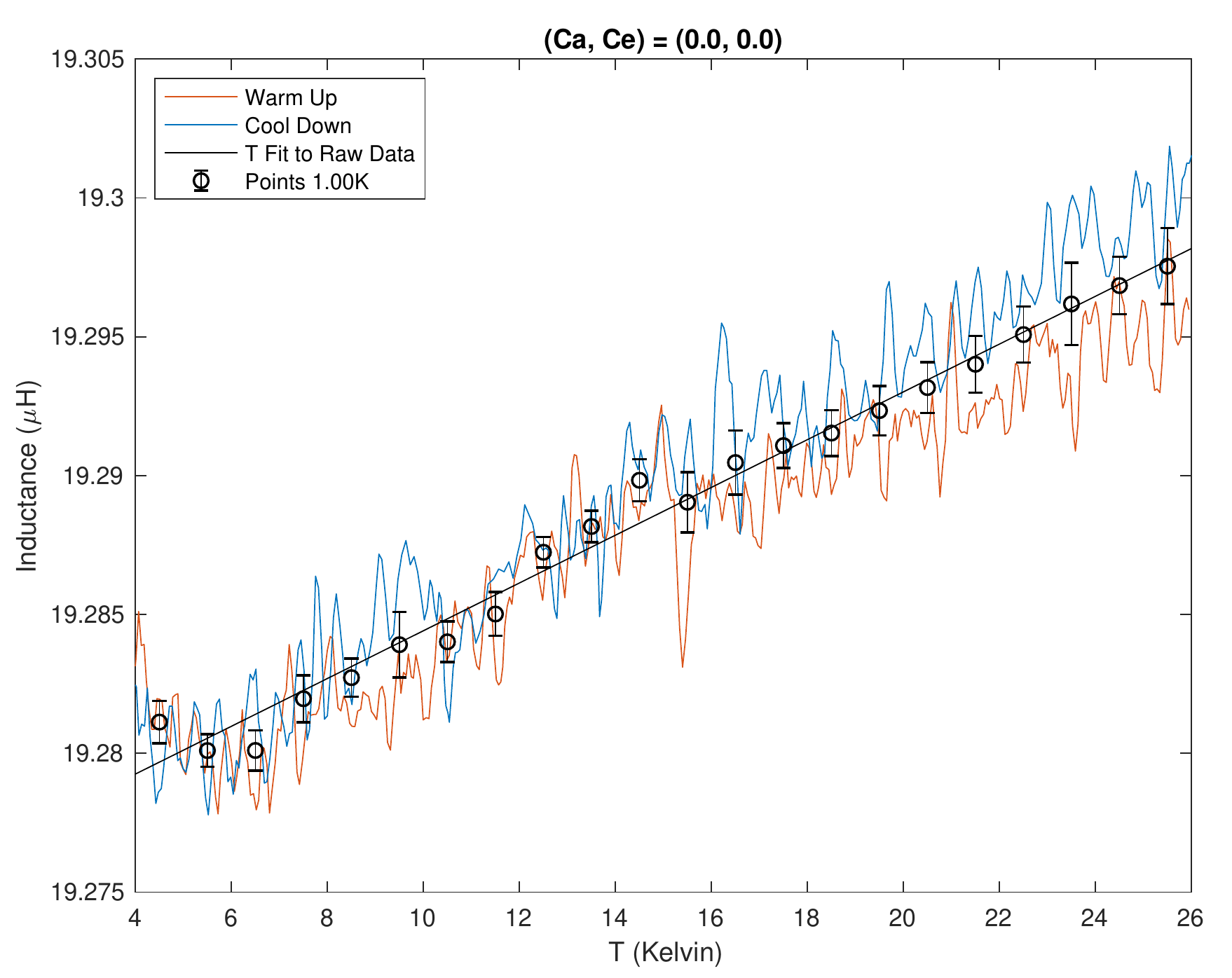}
\caption{
\textbf{Raw data for (0,0) run.}
The blue data is the cool down and the red data is the warm up. The black points are the extracted inductance values, with error bars, obtained from the raw blue and red data as described in the text. The solid line is a linear $T$ fit to the raw data.
}
\label{0000run}
\end{figure}

\begin{figure}[tbp]
\centering \includegraphics[width=100mm]{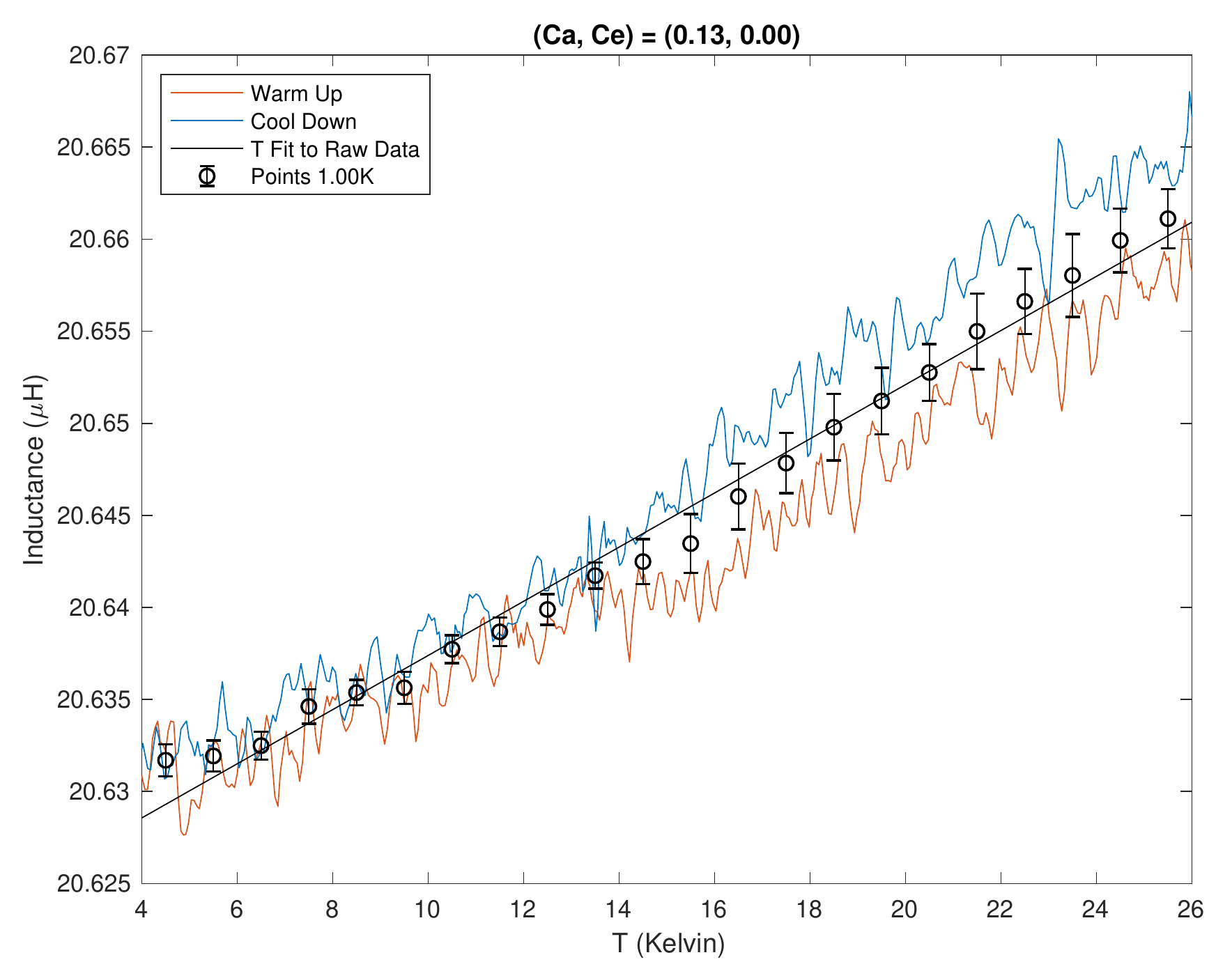}
\caption{
\textbf{Raw data for (13,0) run.}
The blue data is the cool down and the red data is the warm up. The black points are the extracted inductance values, with error bars, obtained from the raw blue and red data as described in the text. The solid line is a linear $T$ fit to the raw data.
}
\label{1300run}
\end{figure}

\begin{figure}[tbp]
\centering \includegraphics[width=100mm]{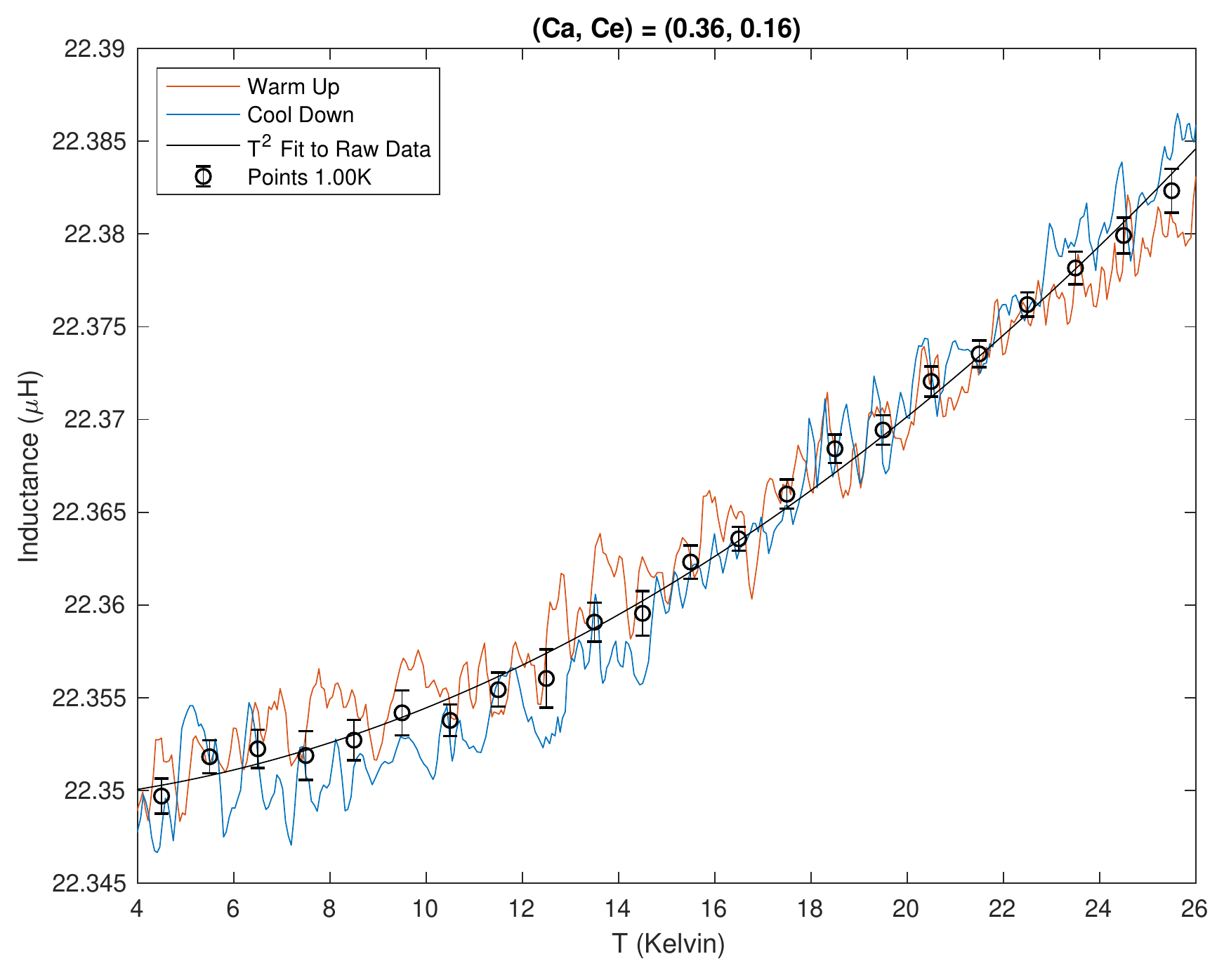}
\caption{
\textbf{Raw data for (36,16) run.}
The blue data is the cool down and the red data is the warm up. The black points are the extracted inductance values, with error bars, obtained from the raw blue and red data as described in the text. The solid curve is a $T^2$ fit to the raw data.
}
\label{3616run}
\end{figure}

\begin{figure}[tbp]
\centering \includegraphics[width=100mm]{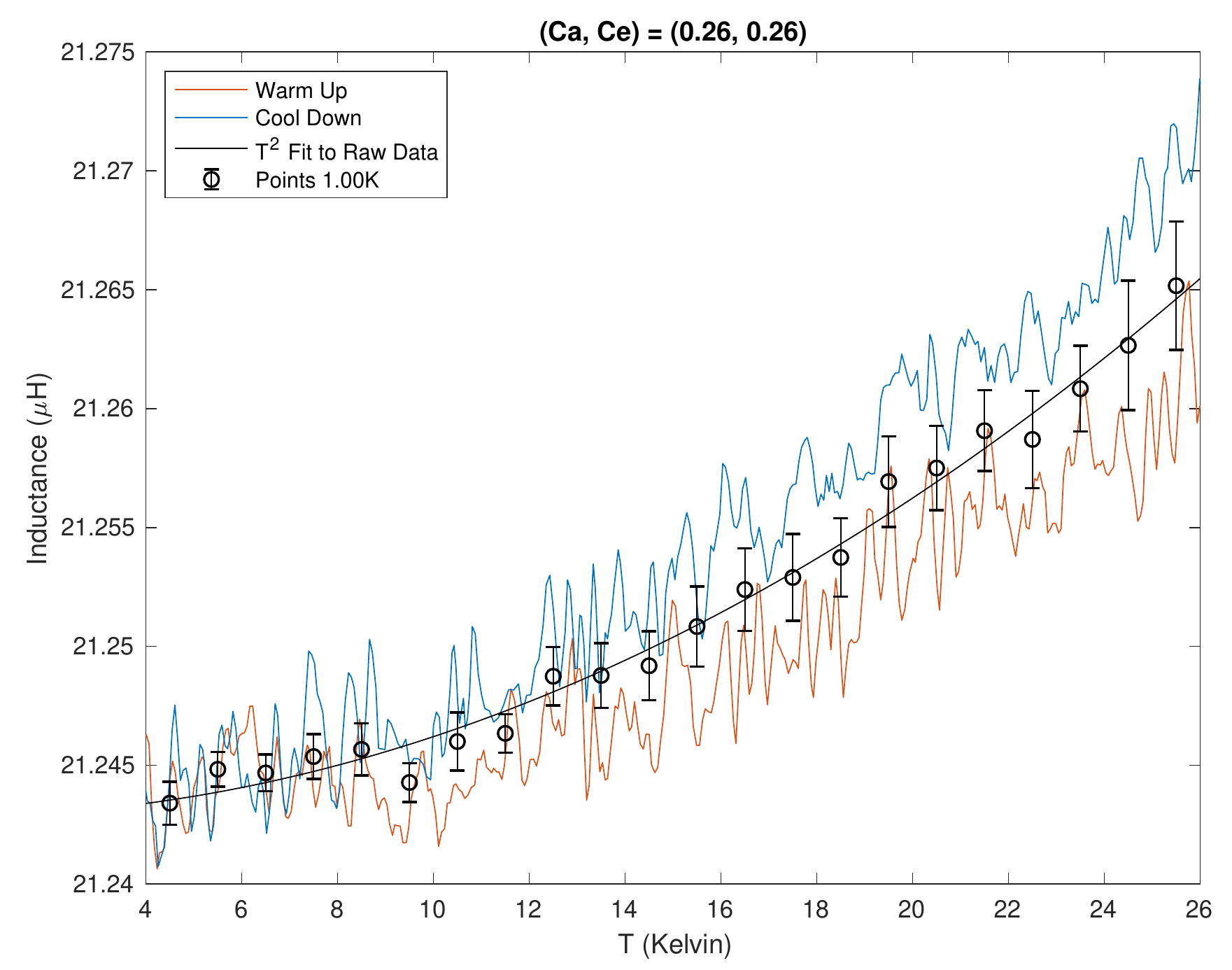}
\caption{
\textbf{Raw data for (26,26) run.}
The blue data is the cool down and the red data is the warm up. The black points are the extracted inductance values, with error bars, obtained from the raw blue and red data as described in the text. The solid curve is a $T^2$ fit to the raw data.
}
\label{2626run}
\end{figure}

\begin{figure}[tbp]
\centering \includegraphics[width=100mm]{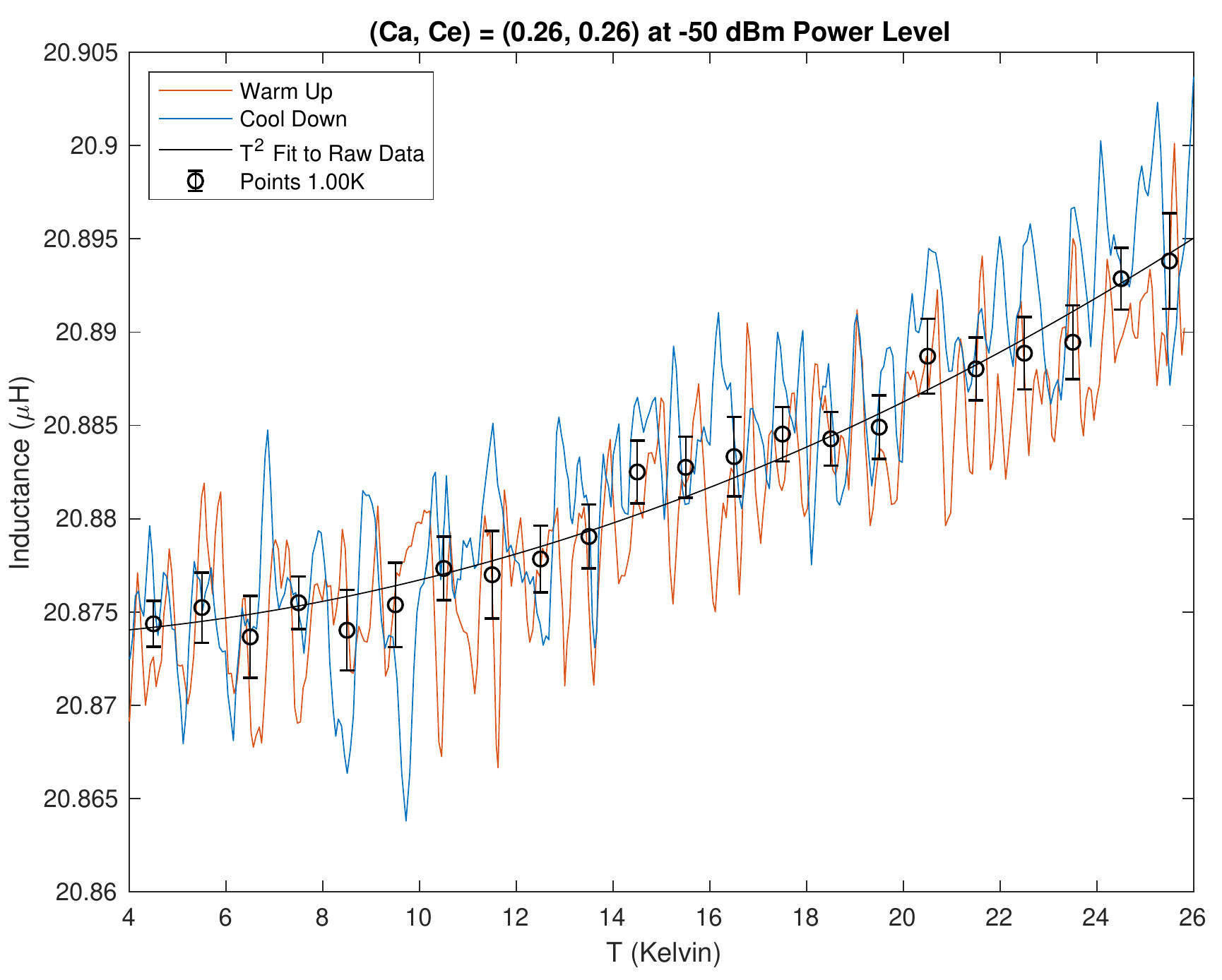}
\caption{
\textbf{Raw data for (26,26) run at $\mathbf{-50\ dBm}$ power.}
The blue data is the cool down and the red data is the warm up. The black points are the extracted inductance values, with error bars, obtained from the raw blue and red data as described in the text. The solid curve is a $T^2$ fit to the raw data.
}
\label{2626-50dBm-run}
\end{figure}

Supplementary Figures~\ref{0000run}, \ref{1300run}, \ref{3616run}, \ref{2626run}, and \ref{2626-50dBm-run} show the raw $L$ for the runs in Figure 4 of the main text. The experiment begins at $26\ \mathrm{K}$ and cools down to $4\ \mathrm{K}$ (blue data) and then warms back up to $26\ \mathrm{K}$ (red data). $840$ data points are acquired during the two temperature sweeps in $\approx 84\ \mathrm{minutes}$ leading to a temperature sweep rate of $\approx 2\ \mathrm{K/min}$ in both directions. Unless otherwise stated, all runs were done at a transmit power level of $-40\ \mathrm{dBm}=0.1\ \mathrm{\mu W}$.

The black curves in the Figures are the best $T$ or $T^2$ fit to the raw data in the blue and red curves. The single data points (black open circles with error bars) are obtained at $1\ \mathrm{K}$ intervals by averaging all the data points (blue and red) within $\pm\ 0.5\ \mathrm{K}$ of the chosen temperature value.
Hence, each data point is an average of $\approx 38$ raw data points.
This averaging removes any hysteresis effects to first order.

The variance of these data points is calcuated. The square root of this variance is an unbiased estimate of the noise error, $\sigma$, in the experiment. The error bars on the black data points is obtained by dividing $\sigma$ by the square root of the number of data points $\approx\sigma/\sqrt{38}$ to obtain the estimated error of the $L$ value of the black data points.

\begin{figure}[tbp]
\centering \includegraphics[width=100mm]{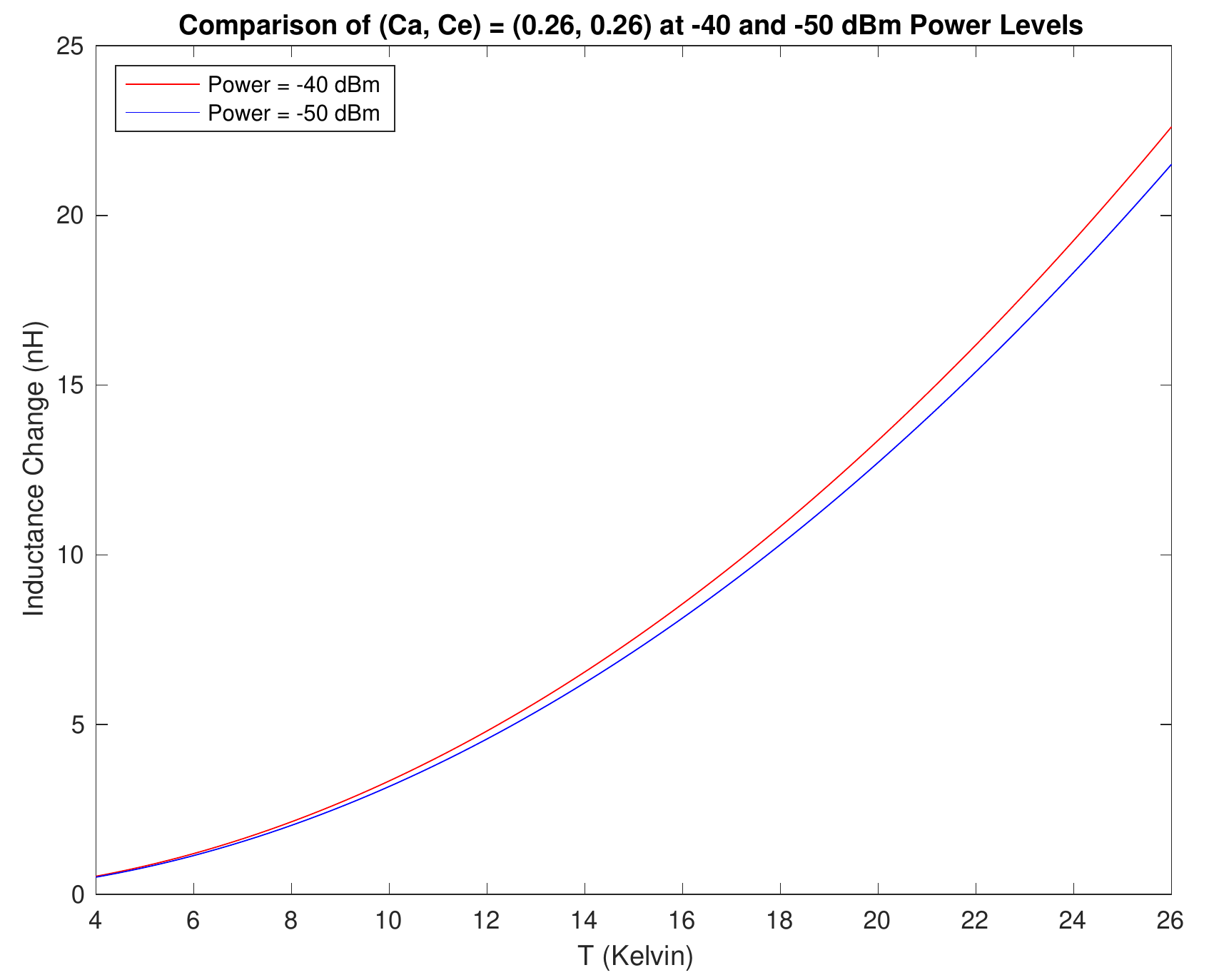}
\caption{
\textbf{Comparison of the (26,26) runs at $\mathbf{-40\ dBm}$ and $\mathbf{-50\ dBm}$ power levels.}
The red curve is the black curve in Supplementary Figure~\ref{2626run} and the blue curve is the black curve from Supplementary Figure~\ref{2626-50dBm-run}. Both curves are plotted with their zero-temperature values subtracted. Within experimental error bars, these two curves are identical.
}
\label{2626-compare}
\end{figure}

Supplementary Figures~\ref{2626run} and \ref{2626-50dBm-run} are two runs of a $(0.26,0.26)$ sample at power levels of $-40\ \mathrm{dBm}=0.1\ \mathrm{\mu W}$ and $-50\ \mathrm{dBm}=0.01\ \mathrm{\mu W}$. Supplementary Figure~\ref{2626-compare} compares the two $T^2$ fits for the change in $L$ for each run. Within the error bars of the experiment, these two runs show there is almost no difference between these two power levels. Hence, all the runs leading to the $T_c$ results in Figure 2 and the inductance change in Figure 4 are performed at $-40\ \mathrm{dBm}$.

\FloatBarrier

\bibliography{supplement.bbl}